\documentclass[11pt,a4paper]{article}

\usepackage[utf8]{inputenc}
\usepackage[T1]{fontenc}
\usepackage[english]{babel}
\usepackage[margin=2.5cm]{geometry}
\usepackage{mathptmx}          
\usepackage{amsmath,amssymb}
\usepackage[version=4]{mhchem} 
\usepackage{graphicx}
\usepackage{float}
\usepackage[format=plain,justification=justified,singlelinecheck=false,
            labelfont=bf,labelsep=period,font=small]{caption}
\usepackage[super,sort&compress,comma]{natbib}   
\usepackage[colorlinks=true,citecolor=blue,linkcolor=blue,urlcolor=blue]{hyperref}
\usepackage{authblk}           

\title{Nanofluid transport in Taylor bubble flows}

\author[1]{Fahad Al-Gburi\thanks{Email: fsa956@student.bham.ac.uk}}
\author[1]{Faysal Khaleel\thanks{Email: fak917@student.bham.ac.uk}}
\author[2]{Yulong Ding\thanks{Email: y.ding@bham.ac.uk}}
\author[1]{Jason Stafford\thanks{Email: j.stafford@bham.ac.uk}}

\affil[1]{\small School of Engineering, University of Birmingham, Birmingham, B15 2TT, United Kingdom.} 
\affil[2]{School of Chemical Engineering, University of Birmingham, Birmingham, B15 2TT, United Kingdom}
\date{}

\begin{document}
\maketitle

\begin{abstract}
\noindent Taylor bubble flows with nanoparticle suspensions are regularly encountered in non-isothermal chemical synthesis and microfluidics. Despite this, a complete mechanistic insight into the coupled thermal and hydrodynamic interactions remains unresolved. Here, a Eulerian–Lagrangian framework has been developed to address this, capturing two‑way momentum, energy exchange, and nanoparticle interactions under non-isothermal conditions. We find that Brownian motion acts in two opposing ways, enhancing local particle-fluid energy exchange yet driving a thermophoretic drift away from the heated wall that strengthens as particle diameter is reduced below 100nm. At particle volume fractions of $\phi \sim 10^{-2}$, Taylor flow departs from the Nusselt scaling with dimensionless slug length, $\tilde{L}_s$, observed for particle-free flows. This coincides with alterations to bubble topology that co-benefit nanoparticle momentum and energy exhange with the carrier fluid. Examining both Poiseuille and Taylor flow problems, we observe contrasting, often opposing dimensionless heat transport behaviours, questioning the suitability of traditionally accepted mixture approaches. This work reveals that, in the visco-inertial regime ($Re > 500$), it is the interplay of hydrodynamic, Brownian, and interfacial forces, combined with the Taylor bubble and slug flow topology, which determine positive or negative $Nu$ outcomes. Interestingly, short-range particle forces at the liquid-gas interface work synergistically with the convective motion in the slug to alter local particle concentration and energy exchanges within downstream thin film and near-wall regions. The intensity of this convective flow was found to impact both slug and film transport phenomena, improving the redistribution of nanoparticles when $\tilde{L}_s \leq 1.6$ and with increasing $Re$.
\end{abstract}

\textbf{Keywords:} Taylor bubble, Poiseuille flow, nanoparticles, microfluidics

\section{Introduction}
\label{sec:headings}

Taylor flow, characterized by alternating elongated gas bubbles and liquid slugs, is widely recognized for its potential to enhance heat and mass transfer in laminar flow systems within mini- and micro-scale devices \citep{Cherlo2010,MAJUMDER2013}. This enhancement arises from the internal circulations generated within bubble/droplet and liquid slug phases, which promote efficient mixing over single phase flows \citep{Taylor1961,oliver1964}. Such flow patterns are commonly observed across a diverse range of applications, including pulsating heat pipes, micro two-phase heat exchangers, digital microfluidics, micro-scale reactor systems, and fuel cells \citep{BETZ2010,KREUTZER2005}. An omnipresent feature across these fields is the inclusion of solid nanoparticles within liquid phases \citep{Abu-Nada2019,Li2024,Ganguly2024,Alrbee2022}. Conventional heat transfer applications often rely on fluids with low intrinsic thermal conductivity \citep{Choi1995}. To overcome this limitation, nanofluids -- created by dispersing nanoparticles in a base fluid -- have emerged as a promising alternative, offering superior thermophysical properties \citep{Jang2007}. Chemical micro-reactors also utilise Taylor flows to synthesise nanoparticles with narrow size distributions compared to batch operations \citep{KREUTZER2005} and which are challenging to account for in hydrodynamic investigations \citep{Angeli2008}. Therefore, understanding the influence of nanoparticles on transport phenomena has emerged as a key research topic, and this work focuses on these effects within Taylor flows.

Early investigations of Taylor flow phenomena were predominantly experimental and theoretical, beginning with foundational studies such as Bretherton's work using a lubrication approximation to investigate thin films ($\delta \ll r)$ with negligible inertia around a confined rising bubble \citep{Bretherton1961}. This provided fundamental insights into bubble formation and flow stability for low Capillary numbers ($Ca \ll 1)$, highlighting critical parameters like bubble shape, speed, and scaling of liquid film thickness ($\delta/r \sim Ca^{2/3}$). Taylor's experiments on viscous liquids extended to $Ca \sim 1$, and an adaptation of Bretherton's scaling law to account for the thickness of the film can be made by replacing $r$ with ($r-\delta$) \citep{Taylor1961,Quere2000}. Using a combination of experiments and scaling arguments, \citet{Quere2000} explored the effect of low viscosity liquids, identifying a visco-inertia regime and showing that non-negligible effects of inertia lead to a significant departure (increase) in film thickness compared to previous relations that depend only on $Ca$, even for low $Ca \sim 0.01$ conditions.

Investigations on the effect of $Re$ have explored further into the visco-inertia regime. \citet{Kreutzer2005a} examined the pressure drop, film thickness and bubble velocity for $Re<900$ using experiments and a 2D finite element method that ignored the gas phase by solving the governing equations within the liquid phase only. \citet{HAN2009} and \citet{WALSH2009} developed empirical relationships for film thickness and pressure drop across a wide range of Reynolds and Capillary numbers ($Ca < 0.2$, $Re < 2000$) and micro-scale and milli-scale capillaries. These studies found that $Ca$, $Re$, and $We$ are the primary scaling parameters that influence film thickness when gravitational effects are negligible ($Bo \ll 1$), while non-dimensional slug length ($\tilde{L}_s = L_s/D$) and $Ca/Re$ describe the contribution of Taylor flow on the total pressure drop in continuous gas-liquid segmented flows. More recently, \citet{Yu2021} examined the effects of Bond number through experiments, lubrication theory and direct numerical simulations (DNS) using the volume of fluid (VoF) method. The results showed that the motion of an elongated bubble in a vertical capillary continues to provide interesting phenomena even for the inertialess regime, where non-unique solutions and breaking of bubble (and film) axisymmetry were observed when external flows are imposed.

Aside from investigations into purely hydrodynamic effects, several studies have implemented both interface capturing VoF  \citep{GUPTA2009,Etminan2023,ZHANG2016-2,Lei2021,Stafford2016} and front-tracking methods \citep{Fischer2010} to resolve both the hydrodynamics and heat transport within the continuous phases of Taylor flows. Experimental measurements of Nusselt number for liquid-gas and liquid-liquid flow configurations have confirmed that slug length and film thickness play a crucial role in heat transfer characteristics and can be described by two asymptotic limits  \citep{Hughmark1965,Etminan2023,Walsh2010,EAIN2015}. In the long slug limit ($L_s^*\gg 1$, where $L_s^* = L_s/DRePr$), the Nusselt number approaches that of single phase Poiseuille flow \citep{Leung2010}. As slug length reduces to $\tilde{L}_s \sim 1$, radial transport is promoted by stronger fluid circulation and $Nu$ approaches that of the theoretical plug flow limit \citep{Walsh2010}. Flow circulations also lead to oscillations in the local $Nu$ profile during the transition from developing to fully developed flow. These show a characteristic period that correlate with one internal circulation and dampen when film thickness or $Ca$ increases \citep{Walsh2010,EAIN2015}.  

The addition of nanoparticles can also have a significant effect on convective heat transfer. The most commonly used particles are aluminium and copper metal oxides, Al$_2$O$_3$ and CuO, with bulk volume fractions typically $\phi < 0.1$. However, many different nanofluids are possible, including emerging multi-functional materials such as h-BN which possesses unique combinations of thermally conductive and electronically insulative properties \citep{Hou2018}. In a recent study by \citet{Alrbee2022} using Al$_2$O$_3$ nanoparticles, $Nu$ was found to increase 35\% for Poiseuille flows and 44\% for Taylor flows with a liquid fraction, $\alpha_l = 0.5$. \citet{Buongiorno2006} formulated a two-component mixture model with the aim of describing convective transport of nanofluids and reconciling differences between experimental observations that showed extraordinarily high heat transfer coefficients and predictions based on traditional one-fluid $Nu$ correlations that included only thermal conductivity and nanoparticle dispersion effects. Instead, the mixture model suggested that particle forces which are dominant at the nanoscale may lead to local variations in nanofluid properties in the boundary layer and this is the primary reason for heat transfer enhancement. 

\begin{figure}
  \centerline{\includegraphics[scale=0.63]{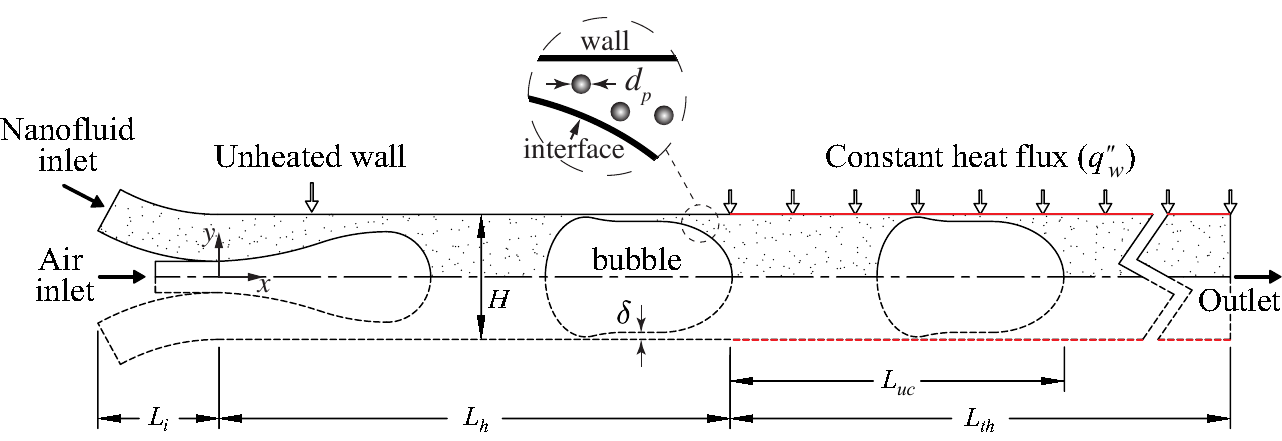}}
  \caption{Schematic showing Taylor bubble formation and transport through a uniformly heated channel section with nanoscale particles present in the continuous liquid phase.}
\label{fig:schematic}
\end{figure}

A challenge in studying hydrodynamic problems involving nanofluids is the disparity of length scales between individual particles and macroscale flow features (e.g. $L/d_p \sim 10^8$, figure \ref{fig:schematic}), and the vast number of particles present in the carrier phase ($N \sim d_p^{-3}$ for constant $\phi$). Previous studies on the effects of nanoparticles have overcome this by assuming homogeneous intrinsic properties and constitutive models to represent an effective viscosity in the Navier-Stokes equations \citep{HinchLeal1975,Buongiorno2006,Alrbee2022}. The commonly used model for nanofluids is the generalised Einstein equation for the viscosity of a fluid containing dilute suspensions, $\mu^*=\mu\left(1+2.5 \phi\right)$, with modifications to the $\phi$ prefactor occasionally introduced to account for empirical observations \citep{Einstein1906,Breki2018}. \citet{Batchelor1977} extended this to include the hydrodynamic effect of Brownian motion on bulk stress when it becomes a dominant factor ($\mu d_{p}^3 E / k T \ll 1$). Although such constitutive models are useful to assess the influence of nanoparticles on bulk stress, the effective viscosity ($\mu^*$) is described exclusively using the suspension phase fraction ($\phi$) and the base fluid viscosity ($\mu$).

The simplifying assumptions that manage effective properties in Eulerian-Eulerian methods assume locally dilute conditions, can neglect important particle forces at these reduced scales (e.g. Brownian, thermophoretic, slip-velocity), and ultimately provide a limited insight to the role of particle migration on nanofluid flow behaviour \citep{BAHIRAEI2016,Abu-Nada2019}. In the case of Taylor flow, and other multiphase flow systems, additional considerations are required to capture interactions with phase interfaces, which for nanoparticles are governed by the interfacial chemistry of the particle-fluid system \citep{Khaleel2025,NguyenSchulze2004}. Furthermore, considering $\delta/r \sim Ca^{2/3}$ when inertia and gravitational effects are negligible in Taylor flows, liquid films are often thick enough to allow nanoparticles to pass through, yet sufficiently thin that an assumption of homogeneous fluid properties is no longer valid (e.g. $\delta \sim 10d_p$ when $Ca\sim 10^{-4} - 10^{-3}$ in micro- and milli-scale capillaries). Indeed, the preferential transport of non-Brownian sub-micron and micron scale particles has been observed numerically and experimentally in isothermal Taylor flows, confirming the importance particle-fluid coupling effects have on the hydrodynamic behaviour \citep{Khaleel2025,Yu2018}. Given that Brownian motion and thermophoretic effects are expected to have a non-negligible influence on nanoscale particle transport ($d_p < 100$ nm), there is a need to consider the coupling between particles, fluid flow and energy to construct a detailed understanding of the hydrodynamics and transport phenomena in Taylor flow suspensions.   

In this work, we address this by developing and implementing an Eulerian-Lagrangian modelling approach that is fully coupled to reveal nanofluid transport in non-isothermal Taylor flows. This approach captures the exchange of momentum and energy between phases by combining DNS using the VoF method to resolve liquid and gas phases, with a Lagrangian method that represents the solid nanoparticle phase. We describe the problem formulation (Sec. \ref{sec:formulation}), numerical techniques implemented (Sec. \ref{sec:implementation}), and the validation and benchmarking of these techniques against previous literature (Sec. \ref{sec:validation}). We then examine how nanoparticle transport and heat transfer are influenced by Brownian diffusion and the resulting thermophoretic drift, particle--interface interactions, particle diameter ($78 \leq d_p \leq 136~\mathrm{nm}$), and particle volume fraction ($10^{-3} \leq \phi \leq 10^{-2}$), together with the principal Taylor-flow parameters of Reynolds number ($500 \leq Re \leq 1000$), Capillary number ($3.45 \times10^{-3}  \leq Ca \leq 6.9 \times10^{-3}$), normalised unit-cell length ($2.66 \leq L_{uc}/H \leq 5$), and void fraction ($0.33 \leq \varepsilon \leq 0.66$) (Sec.~\ref{sec:results}). Particular attention is given to how these particle- and flow-scale parameters govern particle redistribution, bubble and liquid-film structure, particle--fluid thermal exchange, and the resulting local and wall-averaged heat transfer response.

\section{Problem formulation}\label{sec:formulation}

\noindent This section outlines the mathematical and numerical framework employed to model the multi-phase hydrodynamic and heat transfer problem illustrated in figure \ref{fig:schematic}. Here, the liquid carrier and Taylor bubbles (gas) are treated as continuum phases, with the solid nanoparticles modelled as a discrete phase. Our approach to solving this problem is through the integration of Volume of Fluid (VoF) and Multiphase Particle-in-Cell (MPPIC) methods. First, we outline the governing equations used to describe the transport of the continuous phases and to resolve the gas–liquid interfaces in the presence of particles with the VoF method. The second part focuses on the Lagrangian formulation for particle transport and interfacial interactions within the MPPIC framework. Together, these methods form the basis of a coupled Eulerian–Lagrangian approach developed in this work for simulating non-isothermal Taylor flows laden with nanoparticles.

\subsection{Volume of Fluid method} \label{sec:vof-method}
\noindent The gas–liquid interface dynamics in the present work are modeled using a one-fluid Eulerian framework known as the Volume of Fluid (VoF) method \citep{Hirt1981}, which captures the volume fraction of immiscible fluids within a fixed mesh. In the presence of a dispersed solid particle phase, the coupled system is governed by the conservation equations for mass, volume fraction, momentum, and energy:

\begin{equation}\label{Eqn:Mass conservation} \frac{\partial \alpha_c}{\partial t} + \nabla \cdot (\alpha_c \mathbf{u}) = 0 \end{equation}

\begin{equation}\label{Alpha transport}
        \frac{\partial\left(\alpha_c \alpha\right)}{\partial t}+\nabla \cdot \left(\alpha_c \alpha \mathbf{u}\right)+\nabla \cdot \left(\alpha_c \alpha \left(1-\alpha\right) \mathbf{u}_c\right)=0
\end{equation}

\begin{equation}\label{Eqn:Momentum conservation} \frac{\partial (\alpha_c \rho \mathbf{u})}{\partial t} + \nabla \cdot (\alpha_c \rho \mathbf{u} \mathbf{u}) - \left( \frac{\partial \rho}{\partial t} + \nabla \cdot (\rho \mathbf{u}) \right) \mathbf{u} = -\nabla p + \nabla \cdot (\alpha_c \boldsymbol{\tau}) + \alpha_c \rho \mathbf{g} + \mathbf{f}_\sigma + \mathbf{S}_m \end{equation}

\begin{equation}\label{Eqn:Energy conservation} \begin{split} \frac{\partial (\rho c_p T)}{\partial t} + \nabla \cdot (\rho c_p \mathbf{u} T) - \left( \frac{\partial \rho c_p}{\partial t} + \nabla \cdot (\rho c_p \mathbf{u}) \right) T = \nabla \cdot (\kappa \nabla T) + \mathbf{S}_e \end{split} \end{equation}

\noindent where $\mathbf{u}$ describes fluid velocity,  $T$ is the fluid temperature, $\mathbf{g}$ is gravitational acceleration, $p$ is pressure, while the viscous stress tensor, $\boldsymbol{\tau}$, governs momentum dissipation. Within this framework, $\alpha_c$ denotes the volume fraction of the continuous fluid phase, computed as $1 - \alpha_p$ where $\alpha_p$ is the volume fraction of the dispersed particle phase. Meanwhile, $\alpha$ serves as the local phase indicator, taking the value 1 in liquid regions, 0 in gas, and varying continuously between 0 and 1 across the interface. Accordingly, the fluid properties such as density $\rho$, viscosity $\mu$, thermal conductivity $\kappa$, and heat capacity $c_p$ are determined as weighted averages of the liquid and gas phase values:

\begin{equation}\label{density}
    \rho = \alpha \rho_l + (1-\alpha)\rho_g
\end{equation}

\begin{equation}\label{viscosity}
    \mu = \alpha \mu_l + (1-\alpha)\mu_g
\end{equation}

\begin{equation}\label{conductivity}
    \kappa = \alpha \kappa_l + (1-\alpha)\kappa_g
\end{equation}

\begin{equation}\label{heatCapacity}
    c_p = \left[\alpha \rho_l c_{p,l} + (1-\alpha)\rho_g c_{p,g}\right]/{\rho}
\end{equation}

\noindent The thermal conductivity $\kappa_l$ and viscosity $\mu_l$ of the liquid phase (water in this work) were treated as temperature-dependent properties \citep{Touloukian1975,ALNEAMA2018}:

\begin{equation}\label{viscosity_liquid}
\mu_l(T) =  2.414 \times 10^{-5} \; \times \; 10^{\tfrac{247.8}{\,T - 140}}
\end{equation}

\begin{equation}\label{conductivity_liquid}
  \kappa_l(T) = -0.86908 \;+\; 8.9488 \times 10^{-3} T \;-\; 1.5836 \times 10^{-5} T^{2} \;+\; 7.97543 \times 10^{-9} T^{3}
\end{equation}

\noindent This approach is adopted as the liquid phase is in direct contact with the heated wall along the channel, where a significant temperature gradient exists. As a result, the liquid plays a dominant role in the development of the thermal boundary layer adjacent to the wall. Additionally, since the particles are suspended within the liquid phase, it is essential to accurately capture the interactions between the particles and the surrounding fluid. Incorporating temperature-dependent properties enhances model fidelity and allows for a more realistic representation of the coupled hydrodynamic and heat transfer mechanisms involved.

The interphase momentum transfer from dispersed particles is represented by the source term $\mathbf{S}_m$, and interphase thermal coupling is accounted for through the heat source term $\mathbf{S}_e$. In both the momentum and energy equations (eqns. \ref{Eqn:Momentum conservation}, \ref{Eqn:Energy conservation}), the third term on the left-hand side subtracts the spurious implicit diagonal contribution that arises when transient and convective operators are discretized with variable coefficients (product-rule expansion). This cancellation leaves only the physical transport -- \(\alpha_c\rho\,\partial\mathbf{u}/\partial t\) in momentum and \(\rho c_p\,\partial T/\partial t\) in energy -- thereby improving stability and coupling in variable-density or variable-property flows \citep{Jasak1996}. 

The interfacial surface tension $\mathbf{f}_\sigma$ is modelled using the Continuum Surface Force (CSF) approach \citep{BRACKBILL1992}: 

\begin{equation} \mathbf{f}_\sigma = \sigma \aleph \frac{-\rho \nabla \alpha}{0.5(\rho_l + \rho_g)} \end{equation}

\noindent where $\aleph = \nabla \cdot \left( \frac{\nabla \alpha}{|\nabla \alpha|} \right)$ denotes the local interface curvature, computed as the divergence of the unit normal to the interface, and $\sigma$ is the interfacial tension coefficient. To account for its temperature dependence, the interfacial tension was approximated by a linear function over the investigated temperature range \citep{Young1959,Fischer2010}:

\begin{equation}\label{interfacial_tension}
\sigma(T)
=
\sigma_0-\sigma_T\left(T-T_0\right),
\end{equation}

\noindent where $\sigma_0=\sigma(T_0)$ is the interfacial tension at the reference temperature $T_0$, and $\sigma_T=-\left.d\sigma/dT\right|_{T_0}>0$ is the temperature coefficient of interfacial tension.

Together, these formulations (eqns. \ref{Eqn:Mass conservation}- \ref{Eqn:Energy conservation}) provide a consistent framework for resolving multiphase transport in non-isothermal Taylor flows with particle-laden phases, enabling accurate modelling of both momentum and energy exchange across dynamic interfaces.

\subsection{Multi-phase particle-in-cell method}\label{sec:mppic-method}

The MPPIC (Multiphase Particle-in-Cell) method is a computational strategy designed to capture both fluid–solid and solid–solid interactions. As a Lagrangian particle approach, it can resolve the dynamics of spherical particles across regimes ranging from dilute suspensions to dense assemblies, while grouping multiple real particles into a single parcel and giving them the same position, velocity, and acceleration to reduce computational expense for problems involving large numbers of particles. The method was first introduced by \cite{ANDREWS1996} for a one-dimensional configuration, and later generalized to three dimensions by \cite{SNIDER2001}. Within this hybrid framework, interpolation schemes are employed to transfer particle properties from their Lagrangian positions to the Eulerian mesh, where continuum fields such as velocity and volume fraction are computed. The resulting interaction forces are subsequently redistributed to the particles. For high particle concentrations, a particle-stress model is incorporated to approximate collision effects. When coupled with fluid–particle interaction forces, this approach enables a four-way coupled representation of multiphase systems \citep{Kumar2021,CALISKAN2021,Liao2023}.

The particle dynamics within the MPPIC framework are governed by the Liouville equation for the particle distribution function $f(\mathbf{x}_p,\mathbf{u}_p,m_p,t)$, which represents the probability of finding a particle with velocity $\mathbf{u}_p$ and mass $m_p$ at position $\mathbf{x}_p$ and time $t$:

\begin{equation}\label{DF}
    \frac{\partial f}{\partial t} + \nabla_x \cdot (f \mathbf{u}_p) + \nabla_v \cdot (f \mathbf{a}_p) 
    = \frac{f_D - f}{\tau_D} + \frac{f_G - f}{\tau_G}.
\end{equation}

\noindent Here, $\nabla_x$ and $\nabla_v$ denote divergence with respect to the spatial and velocity coordinates, respectively. The first term on the right-hand side of (\ref{DF}) corresponds to the \textit{collision-damping} mechanism originally proposed by \citep{OROURKE2010}, where $f_D$ represents the local mass-averaged velocity distribution and $\tau_D$ is the associated damping timescale. The second term, introduced as the \textit{return-to-isotropy} contribution \citep{OROURKE2012}, relaxes the distribution toward an isotropic equilibrium $f_G$ over a characteristic time $\tau_G$. Detailed formulations for the computation of mean particle fields and discretization strategies for these operators are given in the literature \citep{OROURKE2010,OROURKE2012,OROURKE2014}.

In the present study, the first and second terms on the right-hand side of the distribution equation were omitted. This modelling choice is motivated by two primary considerations. First, the dispersed phase consists of nanoparticles with diameters not exceeding 140~nm, for which collisional damping and return-to-isotropy mechanisms are expected to be insignificant. Second, the particles are suspended in liquid water, whose viscosity is substantially higher than that of gaseous media, resulting in rapid momentum diffusion that further attenuates these effects. In contrast, most studies that explicitly account for these terms have examined micrometre-scale particles ($d_p \sim10-100~\mathrm{\mu m}$) transported in air, such as fluidised bead systems, where their contribution is known to be pronounced \citep{OROURKE2010,OROURKE2012,OROURKE2014}. Under the operating conditions considered here, the influence of these terms is therefore expected to be negligible.

Finally, in equation~(\ref{DF}), the term $\mathbf{a}_p$ denotes the instantaneous acceleration of an individual particle resulting from the combined action of all relevant forces for this Taylor flow problem, and expressed as:

\begin{equation}\label{acceleration1}
    \mathbf{a}_p
    = D_p\left(\mathbf{u}-\mathbf{u}_p\right)
    - \frac{\nabla P}{\rho_p}
    + \mathbf{g}
    - \frac{\nabla \tau_p}{\alpha_p \rho_p}
    +\frac{\left( \mathbf{F}_l
    +\mathbf{F}_v
    + \mathbf{F}_{int}
    + \mathbf{F}_B
    \right)}{m_p}
\end{equation}

\noindent The first four terms constitute the conventional particle acceleration formulation commonly employed in MPPIC-based studies (e.g.~\cite{ANDREWS1996,SNIDER2001,CALISKAN2021}), accounting for the effects of hydrodynamic drag, pressure-gradient force, gravity, and inter-particle stress, respectively. The remaining terms represent additional force contributions arising from lift, virtual mass, gas–liquid interfacial interactions, and Brownian motion. Inclusion of these mechanisms is essential for accurately capturing particle transport and dispersion in nanoscale three-phase flows considered in this study.

The drag coefficient $D_p$ is formulated based on a correlation for the dimensionless drag coefficient $C_D$. In this work, a blended drag model is adopted that combines the Ergun correlation \citep{ergun1952}, applicable to dense particulate regimes, with the Wen--Yu correlation \citep{wen1966}, which is suitable for dilute suspensions. The transition between these two formulations is governed by the local particle volume fraction, following the approach proposed by Gidaspow \citep{gidaspow1994} and subsequently employed in MPPIC-based studies \citep{CALISKAN2021}:

    \begin{equation}
        D_p= \begin{cases}150 \frac{\mu \alpha_p}{\rho_p \alpha_c^2 d_p^2}+1.75 \frac{\rho\left|u-u_p\right|}{\alpha_c \rho_p d_p}, & \alpha_c<0.8 \\ C_D \frac{3}{4} \frac{\alpha_c \rho\left|u-u_p\right|}{\alpha_c \rho_p d_p} \alpha_c^{-2.65}, & \alpha_c \geqslant 0.8\end{cases}
    \end{equation}

    \begin{equation}
        C_D=\left\{\begin{array}{ll}
        \frac{24\left(1+0.15 (\alpha_c Re_p)^{0.687}\right)}{\alpha_c Re_p}, & \alpha_c Re_p<1000 \\
        0.44, & \alpha_c Re_p \geqslant 1000
        \end{array}\right.
    \end{equation}

\noindent where $Re_p$ denotes the particle Reynolds number, defined as:

\begin{equation}\label{Re_p}
    Re_p = \frac{\rho\, d_p\, \lvert \mathbf{u} - \mathbf{u}_p \rvert}{\mu}.
\end{equation}

\noindent In this expression, $\rho$ and $\mu$ are the density and dynamic viscosity of the continuous phase, respectively, while $d_p$ represents the particle diameter. The pressure-gradient and gravitational contributions in equation~(\ref{acceleration1}) can be reformulated by decomposing the pressure-gradient term into two physically distinct components. One component accounts for the local fluid acceleration evaluated at the particle location, while the other corresponds to the buoyancy effect. This decomposition can be expressed as:

\begin{equation}
    -\frac{\nabla P}{\rho_p}
    =
    \left(
    \left[\frac{D \mathbf{u}}{D t}\right]_p
    - \mathbf{g}
    \right)
    \frac{\rho}{\rho_p},
\end{equation}

\noindent where $\left[D\mathbf{u}/Dt\right]_p$ denotes the material acceleration of the carrier phase evaluated at the particle position. Particle collisions are represented by a particle-phase stress to describe the particle-particle interactions without explicitly resolving individual particle collisions. The particle-phase stress $\tau_p$, which represents the averaged collisional interaction among particles, is modelled using the formulation proposed by \cite{harris1994}:

\begin{equation}\label{eq:tau_p}
    \tau_p = \frac{P_s \alpha_p^{\beta}}
    {\max\!\left(\alpha_{\max} - \alpha_p,\;\epsilon\left(1-\alpha_p\right)\right)}.
\end{equation}

\noindent Here, $P_s$ denotes the particle pressure coefficient, $\alpha_{\max}$ represents the local maximum packing fraction, and $\epsilon$ is a small regularisation parameter introduced to avoid singular behaviour as the solid volume fraction approaches close packing. The exponent $\beta$ governs the sensitivity of the particle stress to local concentration, such that increasing $\alpha_p$ produces a rapid increase in $\tau_p$. As discussed by Vargas \emph{et al.}~\citep{Vargas2024}, the appropriate magnitude of the particle-stress parameters $P_s$ and $\beta$ depends on many factors, especially the viscosity of the carrier phase and the characteristic particle size. Consistent with these findings, larger values of $P_s$ and $\beta$, together with smaller $\epsilon$, intensify the effective solid stress and particle interactions. In the present work, $P_s$ and $\beta$ were selected through systematic numerical testing, yielding values suited to a viscous liquid carrier and nanoscale particles. Additional details on this calibration procedure are provided in Sec.~\ref{sec:implementation}.

The lift force acting on the particles, denoted by $\mathbf{F}_l$, is modelled using an empirical formulation originally derived by \cite{saffman1965} and subsequently refined by \cite{MEI1992}. In this approach, the lift force is expressed as:

\begin{equation}
    \mathbf{F}_l = C_l \, \mathbf{F}_{l(\mathrm{Saff})},
\end{equation}

\noindent where $\mathbf{F}_{l(\mathrm{Saff})}$ represents the classical Saffman lift force, given by \cite{crowe2011}:

\begin{equation}\label{saff}
    \mathbf{F}_{l(\mathrm{Saff})}
    =
    \frac{1.615\,(\rho \mu)^{1/2}\, d_p^{\,2}}
    {|\boldsymbol{\omega}|^{1/2}}
    \left(\mathbf{u}-\mathbf{u}_p\right)\times\boldsymbol{\omega}.
\end{equation}

\noindent The validity of equation~(\ref{saff}) is restricted to low-Reynolds-number regimes, subject to the conditions:

\begin{equation}
\begin{array}{ll}
    Re_s=\dfrac{|\mathbf{u}-\mathbf{u}_p|\,d_p}{\nu} \ll 1, 
    & 
    Re_\Omega=\dfrac{\Omega\, d_p^{\,2}}{\nu} \ll 1,\\[6pt]
    Re_G=\dfrac{\dot{\gamma}\, d_p^{\,2}}{\nu} \ll 1,
    &
    \varepsilon=\dfrac{Re_G^{1/2}}{Re_s} \gg 1,
\end{array}
\end{equation}

\noindent where $\boldsymbol{\omega}$ is the local fluid vorticity evaluated at the particle centre, defined as $\boldsymbol{\omega}=\nabla\times\mathbf{u}$. The particle rotational velocity $\boldsymbol{\Omega}$ is estimated as $\boldsymbol{\Omega}=\boldsymbol{\omega} d_p / \left(2|\mathbf{u}-\mathbf{u}_p|\right)$, and $\dot{\gamma}$ denotes the local shear rate, given by $d\mathbf{u}/dt$. The lift coefficient $C_l$ can be expressed as:

    \begin{equation}
        C_l= \begin{cases}(1-\alpha_s) exp(-0.1 Re_s)+\alpha_s, & Re_s \leq 40 \\ 0.0524 (\beta_s Re_s)^{1/2}, & Re_s > 40\end{cases}
    \end{equation}
    
\noindent where $\alpha_s= 0.3314 {\beta_s}^{1/2}$, and $\beta_s$ is as follows:
    \begin{equation}
        \beta_s=\frac{\dot{\gamma} d_p}{2\left|\mathbf{u}-\mathbf{u}_p\right|}=\frac{Re_s \varepsilon^2}{2}=\frac{Re_G}{2 Re_s}
    \end{equation}

\noindent When a particle accelerates relative to the surrounding fluid, it must also accelerate a portion of the adjacent fluid, giving rise to the so-called virtual (or added) mass force. For a spherical particle, the virtual mass force is commonly expressed as \citep{Maxey1983}:

\begin{equation}
\mathbf{F}_{v} = C_{v}\,m_p \left( \frac{D\mathbf{u}}{Dt} - \frac{d\mathbf{u}_p}{dt} \right),
\end{equation}

\noindent where $C_{vm}=0.5$ for a rigid sphere. For the present nanoparticle suspensions, the particle Stokes number is much smaller than unity, implying that particle and fluid accelerations remain closely coupled and the magnitude of the virtual mass force is limited \citep{Maxey1983,Auton1988}. Nevertheless, this term is retained to ensure physical consistency during transient acceleration and deceleration phases, particularly in regions of strong velocity gradients and interfacial flow structures.

To account for the force exerted on particles by the gas--liquid interface, a continuum-based interfacial force model is employed. This force is expressed in the compact form:

\begin{equation}\label{eq:F_int/Final}
    \mathbf{F}_{int} = - C_{int}\, m_p \nabla \alpha ,
\end{equation}

\noindent where $\nabla\alpha$ is the local gradient of the phase-indicator field and $C_{int}$ is an interface force coefficient. The work of \citet{Khaleel2025} and \citet{WASHINO2013} relate this coefficient to the capillary force at the three-phase contact line of a particle that straddles the interface, which is appropriate for non-Brownian, micrometre-scale particles considered there. For the nanoscale particles considered in the present study, a different representation is required, as three-phase contact is not established. Bare Al$_2$O$_3$ is hydrophilic and does not attach spontaneously to air-water interfaces; foam stabilisation with Al$_2$O$_3$ particles of 70--200~nm requires in-situ hydrophobisation to contact angles of about $60^{\circ}$ \citep{Gonzenbach2006}. The approach of a hydrophilic particle towards a bubble through water is instead governed by the surface forces of extended DLVO theory \citep{NguyenSchulze2004}. The total interaction energy and force on approach to an interface ($h\rightarrow0$) can be considered as a combination of van der Waals, electric double-layer, and hydration effects: 

\begin{equation}
  W(h) = W_{\mathrm{vdW}}(h) + W_{\mathrm{EDL}}(h) + W_{\mathrm{hyd}}(h), \qquad
  F(h) = -\frac{\mathrm{d}W}{\mathrm{d}h},
  \label{eq:net_int_F}
\end{equation}

\noindent The net force on a particle approaching the interface becomes $\mathbf{F}_{int} = F(h)\,\mathbf{n}$, with $\mathbf{n}$ the unit normal to the interface directed from the gas into the liquid. 

For Al$_2$O$_3$-water nanofluids considered in this work, the van der Waals interaction for a solid-water-air system has a negative effective Hamaker constant ($A_{132} \approx -4\times10^{-20}$~J, \citealp{Israelachvili2011,Bergstrom1997}) and is therefore repulsive, as measured directly between bubbles and hydrophilic solids \citep{Fielden1996,Englert2009,Tabor2011}. The electric double-layer interaction is influenced by pH: the air-water interface is negatively charged above pH~$\approx$~2-4.5 \citep{Takahashi2005, Beattie2009}, whereas Al$_2$O$_3$ is positively charged below its isoelectric point at pH~$\approx$~8-9.5 \citep{Kosmulski2021}. Hence, the interaction is attractive in acid-stabilised suspensions and repulsive above the isoelectric point, with a range set by the Debye length. The hydration repulsion of hydrated oxide surfaces decays over about 1~nm and prevents contact irrespective of pH \citep{DerjaguinChuraev1974,IsraelachviliPashley1983}. The hydrophobic attraction that drives bubble-particle attachment in flotation \citep{YoonMao1996} is absent for hydrophilic Al$_2$O$_3$ considered in this work. 

The net interaction is thus a short-range repulsion that confines the particles to the liquid phase, modulated in strength and range by the pH and ionic strength of the suspension (Appendix \ref{appB}). Equation \ref{eq:F_int/Final} is used here as a lumped representation of this net repulsion: the force is confined to the interface through $\nabla\alpha$ and directed from the gas towards the liquid. The interface force coefficient, $C_{int}$, is treated as a measure of the strength of the particle-interface repulsion, set by the interfacial chemistry of the suspension. Using this interpretation, varying $C_{int}$ with a fixed particle material and size corresponds to varying the interfacial chemistry of the suspension, which is most commonly achieved in experiments by adjusting pH. The numerical treatment of the force is described in Sec. \ref{sec:Interface-force correction} and Appendix \ref{appB}, and the influence of $C_{int}$ on the particle distribution and energy transfer is examined in Sec. \ref{sec:interface_coefficient}.

For nanoparticles suspended in a liquid carrier phase, random collisions with surrounding fluid molecules induce stochastic particle motion, commonly referred to as Brownian motion \citep{Einstein1905}. In Eulerian-Lagrangian particle-tracking models, this effect is commonly represented through a Langevin-type stochastic forcing term \citep{Li1992,Abuzeid1991}. The Brownian force is described as:

\begin{equation}
\mathbf{F}_{B} = \boldsymbol{\xi}
\sqrt{\frac{6\pi \mu_l k_B T d_p}{C_c \Delta t}},
\label{eq:Brownian_force}
\end{equation}

\noindent where $\boldsymbol{\xi}=(\xi_x,\xi_y,\xi_z)$ is a vector of independent Gaussian random numbers with zero mean and unit variance, $k_B$ is the Boltzmann constant, $T$ is the local continuous phase temperature, $d_p$ is the particle diameter, $\mu_l$ is the liquid dynamic viscosity, $\Delta t$ is the particle integration time-step, and $C_c$ is the Cunningham-type slip correction factor retained in the present implementation. For the liquid-phase conditions considered here, $C_c$ remains close to unity.

The interpretation of thermophoretic motion differs between gas-particle and liquid-solid systems. Classical thermophoretic models such as those of Epstein, Brock and Talbot were developed for particles suspended in gases \citep{Epstein1929,Brock1962,Talbot1980}. By contrast, \citet{Michaelides2013,Michaelides2015} argued that, in liquids, thermophoresis should be viewed as the ensemble-averaged drift arising from temperature-dependent Brownian motion. Accordingly, the present formulation retains the stochastic Brownian-force contribution and attributes the resulting non-isothermal drift to thermophoresis, rather than introducing a separate thermophoretic force derived from gas-phase theory. Details of the numerical implementation of this formulation, together with its assessment against the isothermal benchmark of \citet{Li1992} and the non-isothermal benchmark of \citet{Michaelides2015}, are provided in Sec.~\ref{sec:implementation}.

In isothermal conditions, the stochastic forcing represented by equation (\ref{eq:Brownian_force}) reproduces the classical Stokes--Einstein diffusion behaviour. In non-isothermal conditions, however, the Brownian impulse strength becomes spatially dependent through the local temperature $T$, such that particles subjected to stronger thermal agitation in hotter regions and weaker agitation in colder regions acquire a non-zero ensemble-averaged drift. This drift is interpreted as thermophoresis in liquids \citep{Michaelides2015}. Accordingly, in the present formulation, thermophoresis is not introduced as an additional deterministic force term independent of Brownian motion; instead, it emerges naturally from the stochastic Brownian forcing under a temperature gradient.

By taking velocity and mass moments of equation~(\ref{DF}), the Eulerian governing equations for the particle phase can be derived. Specifically, the conservation equations are obtained by multiplying equation~(\ref{DF}) by $\alpha_p m_p$ and $\alpha_p m_p \mathbf{u}_p$, followed by integration over the particle velocity and mass spaces \citep{SNIDER2001,CALISKAN2021}. This procedure leads to the particle-phase continuity equation:

    \begin{equation}\label{p-cont}
        \frac{\partial\left(\overline{\alpha_p \rho}_p\right)}{\partial t}+\nabla \cdot \left(\overline{\alpha_p \rho}_p \overline{\mathbf{u}}_p\right)=0
    \end{equation}

\noindent Similarly, the particle momentum equation is:

\begin{gather}
    \frac{\partial\left(\overline{\alpha_p \rho}_p \overline{\mathbf{u}}_p \right)}{\partial t}+\nabla \cdot \left(\overline{\alpha_p \rho}_p \overline{\mathbf{u}}_p 
    \overline{\mathbf{u}}_p\right)=-\alpha_p \nabla P-\nabla \tau_p
    -\overline{\alpha_p \rho}_p C_{int}\nabla\alpha+\overline{\alpha_p \rho}_p \boldsymbol{g}
    \notag\\
    +\, \overline{\alpha_p \rho}_p\boldsymbol{\xi}\sqrt{\frac{6\pi \mu_l k_B T d_p}{C_c \Delta t}}
     +\overline{\alpha_p \rho}_p\, C_v\left(\frac{D\mathbf{u}}{Dt}-\frac{d\mathbf{u}_p}{dt}\right)
    +\iint f m_p D_p\left(\mathbf{u}-\mathbf{u}_p\right) d\mathbf{u}_p dm_p
    \notag\\
    +\iint f C_l \frac{1.615 (\rho \mu)^{1 / 2} {d_p}^2}{\left| \boldsymbol{\omega}\right|^{1 / 2} }\left(\mathbf{u}-\mathbf{u}_p\right) \times \boldsymbol{\omega} \, d\mathbf{u}_p \, dm_p
    \notag\\
    -\nabla \cdot \left[\iint f m_p\left(\mathbf{u}_p-\overline {\mathbf{u}}_p\right) \left(\mathbf{u}_p-\overline {\mathbf{u}}_p\right) d\mathbf{u}_p dm_p\right]
\label{p-mom}\end{gather}

\noindent In Equations~\eqref{p-cont} and~\eqref{p-mom}, $\overline{\alpha_p \rho}_p$ is the average particle density and is given by:

    \begin{equation}
        \overline{\alpha_p \rho}_p=\iint f m_p dm_p d\mathbf{u}_p 
    \end{equation}

\noindent with the mean particle velocity (mass-average):

    \begin{equation}
        \overline {\mathbf{u}}_p=\frac{1}{\overline{\alpha_p \rho}_p}\iint f m_p\mathbf{u}_p dm_p d\mathbf{u}_p
    \end{equation}

\noindent and the particle volume fraction ($\alpha_p$):
    \begin{equation}
        \alpha_p=\iint f \frac{m_p}{\rho_p} dm_p d\mathbf{u}_p
    \end{equation}

\noindent Finally, the interphase momentum exchange between the particles and the surrounding fluid (source term $\mathbf{S}_m$ in eq. \ref{Eqn:Momentum conservation}) is:

\begin{equation}
    \begin{split}
        \mathbf{S}_m = \iint \left( D_p \left(\mathbf{u} - \mathbf{u}_p\right) - \frac{\nabla P}{\rho_p}+ C_l \frac{1.615 (\rho \mu)^{1 / 2} {d_p}^2}{\left| \boldsymbol{\omega} \right|^{1 / 2}} \left(\mathbf{u} - \mathbf{u}_p\right) \times \boldsymbol{\omega} \right.
        \\
        \left. -C_{int}\nabla\alpha + C_v\left(\frac{D\mathbf{u}}{Dt}-\frac{d\mathbf{u}_p}{dt}\right) +\boldsymbol{\xi}\sqrt{\frac{6\pi \mu_l k_B T d_p}{C_c \Delta t}} 
        \right) f m_p d\mathbf{u}_p \, dm_p
    \end{split}
\end{equation}

\noindent The interphase heat transfer between the dispersed particles and the surrounding fluid is introduced into the continuous phase energy equation (\ref{Eqn:Energy conservation}) through a volumetric source term, $\mathbf{S}_e$. This term represents the cumulative thermal exchange occurring within each computational cell and is evaluated based on the instantaneous thermal state of the particles contained in that cell \citep{Minkowycz2006,BIANCO2009}:

\begin{equation}\label{eq:Se}
\mathbf{S}_e = \frac{1}{\Delta V} \sum_{p=1}^{n_p} m_p c_{p,p} \frac{dT_p}{dt},
\end{equation}

\noindent where $\Delta V$ is the cell volume, $n_p$ is the number of particles located within the cell, $c_{p,p}$ is the particle specific heat capacity, and $T_p$ is the particle surface temperature.

To obtain ${dT_p}/{dt}$, a Lagrangian energy balance is solved for each spherical particle. Assuming uniform internal temperature (lumped-capacitance approximation), the particle energy equation can be written as \citep{Minkowycz2006, Das2008}:

\begin{equation}
m_p c_{p,p} \frac{dT_p}{dt} = h A_p \left( T - T_p \right),
\end{equation}

\noindent where $T$ is the local continuous phase temperature interpolated at the particle position, $A_p$ is the particle surface area, and $h$ is the convective heat transfer coefficient between the particle and the surrounding fluid.

The particle heat transfer coefficient $h$ is evaluated using the Ranz–Marshall correlation \citep{Ranz1952}, which is widely adopted for spherical particles in laminar flows:

\begin{equation}
Nu_p = 2 + 0.6 Re_p^{1/2} Pr^{1/3},
\end{equation}

\noindent where $Nu_p = h d_p / \kappa$ is the particle Nusselt number, and $Pr = \mu c_p / \kappa$ is the Prandtl number of the continuous liquid phase. This formulation ensures consistent coupling between the Lagrangian particle energy equation and the Eulerian thermal field of the carrier phase.

Convective heat transfer along the heated length ($L_{th}$, figure \ref{fig:schematic}) was evaluated from the resolved Eulerian temperature field and applied identically to the Taylor flow and Poiseuille flow cases so that the two flow configurations are comparable. Two complementary measures are used to describe the non-dimensional heat transport. The local Nusselt number resolves the axial structure of the wall response within a bubble/slug unit-cell ($L_{uc}$) and is the quantity through which the film, bubble-cap and slug regions can be distinguished. The averaged Nusselt number, by contrast, characterises the transport behaviour of the entire heated channel section, $L_{th}$.

A uniform wall-normal temperature gradient $\lvert\partial T/\partial n\rvert_w$ was imposed over the heated length, so that the local wall heat flux $q''_w(x)=\kappa_w(x)\lvert\partial T/\partial n\rvert_w$ followed the conductivity of the fluid adjacent to the wall ($\kappa_w$). For each streamwise location the bulk temperature was obtained from the enthalpy-flux-weighted (mixing-cup) definition \citep{Minkowycz2006,TAHIR2012}:

\begin{equation}\label{eq:bulk_temperature}
T_b(x)=
\frac{\displaystyle \int_A T\,\rho_{} c_p\,u_{\parallel}\, dA}
{\displaystyle \int_A \rho_{} c_p\,u_{\parallel}\, dA},
\end{equation}

\noindent where $A$ is the local channel cross-section and $u_{\parallel}=\lvert\mathbf{u}\cdot\hat{\boldsymbol{e}}_f\rvert$ is the velocity component along the prescribed flow direction $\hat{\boldsymbol{e}}_f$, the modulus ensuring that recirculating fluid is weighted correctly in the local average. The local Nusselt number then follows as \citep{BIANCO2009,TAHIR2012}:

\begin{equation}\label{eq:local_nusselt}
Nu_x=\frac{q''_w(x)\,D_h}{\kappa_w(x)\,\Delta T_x}
=\frac{\lvert\partial T/\partial n\rvert_w\,D_h}{\Delta T_x},
\qquad
\Delta T_x=T_w(x)-T_b(x),
\end{equation}

\noindent where $D_h = 2H$ is the characteristic length, with $H$ being the channel height, and $T_{w}$ denoting the local wall temperature.

An arithmetic average of $Nu_x$ does not adequately represent the overall heat transfer in Taylor flows. Within a single unit-cell, $Nu_x$ varies by an order of magnitude, rising sharply in the thin liquid film and falling again in the slug region \citep{Gupta2010,Lei2021}. Since the Nusselt number represents a thermal conductance and its reciprocal a thermal resistance, averaging $Nu_x$ arithmetically returns the mean of $\Delta T_x^{-1}$ rather than the reciprocal of the mean of $\Delta T_x$. The result is dominated by the narrow film maxima, which carry only a small fraction of the wall heat, and does not correspond to the wall-to-bulk thermal resistance of the flow. The averaging is therefore performed on this thermal resistance \citep{Gupta2010}, giving a harmonic rather than an arithmetic mean:

\begin{equation}\label{eq:Nu_bar}
\overline{Nu}
=\left[\frac{1}{L_r}\int_{L_r} Nu_x^{-1}\,dx\right]^{-1}
=\frac{\overline{q''_w}\,D_h}{\kappa_w\left(\overline{T}_w-\overline{T}_b\right)},
\end{equation}

\noindent in which the wall heat flux and the wall and bulk temperatures are averaged separately before the Nusselt number is formed.

\begin{equation}\label{eq:Nu_bar_components}
\overline{q''_w}=\frac{1}{L_r}\int_{L_r} q''_w\,dx,
\qquad
\overline{T}_w=\frac{1}{L_r}\int_{L_r} T_w\,dx,
\qquad
\overline{T}_b=\frac{1}{L_r}\int_{L_r} \hat{T}_b(x)\,dx .
\end{equation}

\noindent where $L_r$ is the length of the region of interest. \citet{Gupta2010} evaluate $\overline{T}_b$ by integrating the mixing-cup temperature over a complete unit-cell and generalise the resulting unit-cell value to the required number of cells, which presupposes a repeating, fully developed train. As the present channel is short and lies within the thermal development region, successive unit-cells are not equivalent, no single cell is representative, and the length that an integer number of them cover is itself parameter-dependent. To overcome this, the bulk temperature averaging in equation (\ref{eq:Nu_bar_components}) is obtained instead from a one-dimensional energy balance on the stream \citep{Shah1978}:

\begin{equation}\label{eq:energy_balance}
\hat{T}_b(x)=T_{in}+\frac{1}{\dot{C}}\int_{x_0}^{x} q''_w(x')\,dx',
\qquad
\dot{C}=\left\langle\int_A \rho c_p\,u_x\,dA\right\rangle,
\end{equation}

\noindent where $T_{in}$ is the inlet temperature, $x_0$ the start of the heated length, $u_x$ the signed axial velocity and $\dot{C}$ the capacity rate of the stream. The latter is evaluated as a single value for each case as $\int_A \rho c_p\,u_x\,dA$ reduces by $\mathcal{O}(10^{-3})$ wherever liquid is displaced by a Taylor bubble, and a local value would reflect the phase occupying the cross-section rather than the throughput of the channel. Equation~\eqref{eq:energy_balance} states that all heat crossing the wall between $x_0$ and $x$ appears as enthalpy carried by the stream at $x$, and reduces in the Poiseuille flow cases to the classical one-dimensional balance, so that one expression serves both flow regimes. The two bulk temperatures thus serve distinct purposes: eq. \eqref{eq:bulk_temperature} resolves the axial structure within a unit-cell and underlies the local profiles, while eq. \eqref{eq:energy_balance} supplies the throughput temperature required for a meaningful section average. All averages are taken over a region of interest, $L_r$, spanning $8.1\le x/H\le 19$. This excludes a short initial section of the heated length ($x-x_0/H = 0.1$), over which $\Delta T_x$ grows from zero, and the $x/H > 19$ channel region, to avoid effects from the outflow condition (located at $x/H = 20$). A temporal average is also required for Taylor flows, since the bubble train translates the channel creating a time-varying local Nusselt number.

Taylor flow is characterised by a continuous liquid film that completely separates the gas bubbles from the channel walls, so that the solid boundary is wetted exclusively by the nanofluid phase. As a result, the wall conductivity, $\kappa_w$, in eq. \eqref{eq:local_nusselt} and the mixture properties in eq. \eqref{eq:bulk_temperature} and eq. \eqref{eq:energy_balance} are those of the local nanofluid rather than a globally averaged mixture. Although particle energy and momentum exchanges were resolved using the Eulerian-Lagrangian approach, the evaluation of the non-dimensional heat transfer ($Nu$, eq. \ref{eq:local_nusselt}, eq. \ref{eq:Nu_bar}) required continuum thermophysical fields defined on the Eulerian mesh. An effective liquid-particle mixture was first constructed within the liquid phase using the local continuous fluid fraction \(\alpha_c=1-\alpha_p\). The corresponding density and heat capacity were defined as \citep{Eastman1999,BIANCO2009}:

\begin{equation}
\rho_{nf}=\alpha_c \rho_l + (1-\alpha_c)\rho_p,
\qquad
c_{p,nf}=
\frac{\alpha_c \rho_l c_{p,l} + (1-\alpha_c)\rho_p c_{p,p}}{\rho_{nf}},
\end{equation}

\noindent while the effective thermal conductivity of the liquid-particle mixture was evaluated using the classical Maxwell model for a dilute suspension of spherical particles dispersed in a continuous liquid phase \citep{Maxwell1873,Fan2011}:

\begin{equation}
\label{thermal_conductivity_Maxwell}
\kappa_{nf}=
\kappa_l
\frac{
\kappa_p+2\kappa_l
+2(1-\alpha_c)(\kappa_p-\kappa_l)
}{
\kappa_p+2\kappa_l
-(1-\alpha_c)(\kappa_p-\kappa_l)
},
\end{equation}

\noindent This effective liquid-particle phase was then combined with the gas phase through the phase indicator \(\alpha\) to construct the fields used to determine the non-dimensional heat transfer:

\begin{equation}
\rho=\alpha \rho_{nf} + (1-\alpha)\rho_g,
\end{equation}
\begin{equation}
c_{p}=
\frac{\alpha \rho_{nf} c_{p,nf} + (1-\alpha)\rho_g c_{p,g}}
{\rho}
\end{equation}

\noindent Using this Eulerian-Lagrangian formulation for three-phase flows containing nanoparticle suspensions, this study explores the coupled, multi-scale hydrodynamic and heat transport mechanisms of Brownian, nanoscale particles within Taylor flows. Table \ref{tab:simulation-parameters} summarises the physical properties of the liquid (water), gas (air), and solid particle (Al$_2$O$_3$) phases that will be used in this work. A schematic of the computational domain is shown in figure \ref{fig:schematic}, illustrating the physical model for the investigation, the co-flow formation and spatial arrangement of Taylor bubbles, and the accompanying liquid-solid slurry.

In addition to the dimensionless heat transfer, eq. (\ref{eq:local_nusselt} and eq. (\ref{eq:Nu_bar}), we introduce the following dimensionless parameters for describing the velocity, temperature, and particle-fluid energy exchanges:  

\begin{equation}
    |\tilde{\mathbf{u}}| = \frac{\rho |\mathbf{u}|D_h}{\mu Re},
\end{equation}
\begin{equation}
    \tilde{\theta} = \frac{(T-T_{in})\kappa}{q_w''D_h},
\end{equation}
\begin{equation}
    \tilde{S}_e =\frac{|\mathbf{S}_e|D_h}{4q''_w}
\end{equation}

\noindent The latter is a dimensionless representation of the Eulerian interphase energy source (eq. \ref{eq:Se}), and describes the relative strength of local particle-fluid energy exchanges to the total energy density of the heated channel.

\begin{table}
  \begin{center}
  \def~{\hphantom{0}}
  \begin{tabular}{lc}
      Parameter (unit) & Value \\[3pt]
      Liquid density, $\rho_l$ ($\mathrm{kg\,m^{-3}}$)
      & 1000 \\
      Liquid viscosity, $\mu_l$ ($\mathrm{Pa\,s}$)
      & $1\times10^{-3}$ \\
      Liquid thermal conductivity, $\kappa_l$ ($\mathrm{W\,m^{-1}\,K^{-1}}$)
      & $0.6$ \\
      Liquid specific heat, $c_{p,l}$ ($\mathrm{J\,kg^{-1}\,K^{-1}}$)
      & $4200$ \\
      Gas density, $\rho_g$ ($\mathrm{kg\,m^{-3}}$)
      & 1.205 \\
      Gas viscosity, $\mu_g$ ($\mathrm{Pa\,s}$)
      & $1.8\times10^{-5}$ \\
      Gas thermal conductivity, $\kappa_g$ ($\mathrm{W\,m^{-1}\,K^{-1}}$)
      & $0.03$ \\
      Gas specific heat, $c_{p,g}$ ($\mathrm{J\,kg^{-1}\,K^{-1}}$)
      & $1000$ \\
      Particle diameter, $d_p$ ($\mathrm{nm}$)
      & 78--136 \\
      Particle density, $\rho_p$ ($\mathrm{kg\,m^{-3}}$)
      & 3950 \\
      Particle thermal conductivity, $\kappa_p$ ($\mathrm{W\,m^{-1}\,K^{-1}}$)
      & $42$ \\
      Particle specific heat, $c_{p,p}$ ($\mathrm{J\,kg^{-1}\,K^{-1}}$)
      & $770$ \\
  \end{tabular}
  \caption{Properties of the liquid, gas, and solid phases used in the present study.}
  \label{tab:simulation-parameters}
  \end{center}
\end{table}

\section{Numerical implementation and validation} \label{sec:implementation}

The boundary conditions were defined to reproduce the generation and subsequent transport of gas-liquid Taylor flow in a two-dimensional plane channel shown in figure~\ref{fig:schematic}. The computational domain represents one half of the channel, with the channel centreline treated as a symmetry plane. The nanofluid was introduced through a single curved inlet located between the central air inlet and the channel wall, while air was injected along the channel centreline through the central inlet. This configuration represents a planar co-flow arrangement, in which the nanofluid stream surrounds the central air stream.

Uniform and constant velocity boundary conditions were prescribed at both the air and nanofluid inlets. A zero-gradient condition was applied to the pressure at all inlet boundaries, allowing the pressure field to develop without imposing an artificial inlet pressure. At the channel outlet, the pressure was fixed at zero as a reference condition, while a zero-gradient condition was imposed on the velocity so that the gas-liquid flow could leave the computational domain without additional constraints. The channel wall was treated using a no-slip velocity condition together with a zero-gradient pressure boundary condition, while a symmetry boundary condition was imposed along the channel centreline.

An unheated hydrodynamic development section of length, \(L_h\), was included upstream of the heated section. This region was provided to allow the bubble-generation process to be completed and the resulting Taylor flow to stabilise before entering the heated channel section. The wall of the hydrodynamic development section was therefore treated as isothermal and maintained at the same temperature as the inlet fluids. Downstream of this region, a constant heat flux was imposed on the outer channel wall over a heated length, \(L_{th}\). The use of a separate unheated development section prevented the applied thermal boundary condition from influencing the initial bubble-formation process.

For the dispersed particle phase, nanoparticle injection was not activated during the first computational time step. Particle loading and injection were initiated at $t_{inj}=9\times10^{-4}~\mathrm{s}$. This delay established the Eulerian gas-liquid velocity and pressure fields before activation of the Eulerian-Lagrangian coupling. At $t=t_{inj}$, particles were loaded homogeneously throughout the liquid phase already present within the computational domain. Thereafter, nanoparticles were introduced continuously with the incoming liquid phase through the nanofluid inlet (figure \ref{fig:schematic}).

The solver was developed within the OpenFOAM-v2312 finite volume computational mechanics framework. Details on the Eulerian-Lagrangian coupling, numerical schemes, solution methods and convergence criteria are provided in Appendix \ref{appA}. To ensure that the numerical methods were suitable for simulating the three-phase flow and heat transfer problem, the key modelling components controlling particle transport and particle-interface interaction were examined systematically beforehand. Particular attention was given to the interface force treatment (eq. \ref{eq:Fint_alphaH}), Brownian force implementation (eq. \ref{eq:Brownian_force}), particle stress model (eq. \ref{eq:tau_p}), and the number of physical particles represented by each computational parcel. These components are directly linked to particle accumulation, particle migration, particle interaction near the gas-liquid interface, and the stability of Eulerian-Lagrangian coupling. Therefore, the following subsections describe the corrective, verification, and validation steps adopted to establish a physically consistent numerical framework for the present study.

\subsection{Interface force correction}
\label{sec:Interface-force correction}

The implementation of the interface force (eq. \ref{eq:F_int/Final}) follows a correction procedure previously introduced by \citet{Khaleel2025}. In that work, the traditional treatment of the particle interface force was found to produce non-physical force contributions away from the gas-liquid interface and, under certain conditions, an incorrect force direction near the interface. To avoid repeating the details here on our previous work, only the correction adopted in the present study is summarised. In the modified formulation, the original phase-fraction field used in the interface force calculation was replaced by a Heaviside-type interfacial field, $\alpha_H$, constructed from the gas-liquid volume fraction field, $\alpha$. The interface force was then evaluated as:

\begin{equation}
    \mathbf{F}_{int} = - C_{int} m_p \nabla \alpha_H ,
    \label{eq:Fint_alphaH}
\end{equation}

\noindent instead of using the gradient of the original volume fraction expression. This treatment confines the force action to the gas-liquid interface and prevents local variations in particle concentration within the liquid region from generating artificial interfacial forces. This agrees with the short-range repulsive interface forces present in this problem, which have been predicted to act at interface distances, $h<120~\mathrm{nm}$, using a simple model to describe the net contributions of van der Waals, electric double-layer, and hydration effects (eq. \ref{eq:net_int_F}, Appendix \ref{appB}). In addition, using $\nabla \alpha_H$ ensures that the force direction remains from the gas phase towards the liquid suspension, thereby reducing the possibility of non-physical particle penetration into the gas phase. This correction is particularly important in the present work on nanofluids, as the predicted nanoparticle distribution near the bubble interface directly affects particle migration, local suspension structure, and the thermal interactions throughout the liquid slug and thin film regions. A baseline value of $C_{int} = 10$ was chosen in this work, reflecting the hydrophilic characteristics of $\mathrm{Al_2O_3}$ nanoparticles. The more broader influence of the interface force coefficient on particle behaviour near the gas-liquid interface, and on the predicted heat transport phenomena, are discussed in Sec. \ref{sec:results}. These coefficients have been chosen to provide a range of particle-interface forces, $0.38~\mathrm{pN} \leq\mathbf{F}_{int}\leq 7.5~\mathrm{pN}$, that are representative of the physical interactions in this aqueous nanofluid system (Appendix \ref{appB}). 

\subsection{Brownian force implementation and validation}
\label{sec:brownian_validation}

The Brownian force described by equation (\ref{eq:Brownian_force}) was evaluated using the local liquid temperature and viscosity interpolated at the particle position, so that the stochastic forcing remained fully coupled to the resolved temperature field. This Brownian force contribution was applied as a three-dimensional Gaussian stochastic vector. This approach recovered the correct second-order statistics of particle dispersion in isothermal conditions and ensured that the non-isothermal drift arising from gradients in the temperature field were represented consistently.

\begin{figure}
    \centering
    \includegraphics[width=1.0\textwidth]{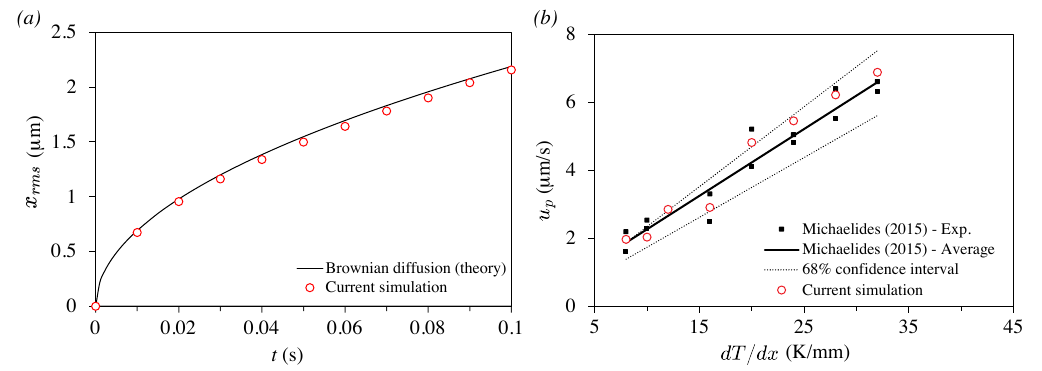}
    \caption{Validation of the Brownian force implementation. (\textit{a}) Predicted root-mean-square particle displacement, $x_{rms}$, of the current model against theory for isothermal Brownian diffusion. (\textit{b}) Predicted particle drift velocity, $u_{p}$, under non-isothermal conditions and compared with the experiments from \cite{Michaelides2015}.}
    \label{fig:validation_brownian_all}
\end{figure}

The model was first verified under isothermal conditions against the theoretical Brownian diffusion behaviour used in stochastic particle transport studies such as \citet{Li1992}. A simple quiescent liquid test case was constructed in which 50 nm particles were released from a localized initial position in a uniform temperature field, and the root-mean-square displacement normal to the mean transport direction ($x_{rms}$) was evaluated as a function of time. The numerical results were then compared with the corresponding theoretical Brownian diffusion relation. As shown in figure \ref{fig:validation_brownian_all}\textit{(a)}, the predictions are in very good agreement with the theoretical curve, confirming that the implementation reproduces the expected isothermal Brownian diffusion behaviour.

The same model was then assessed under non-isothermal conditions against the liquid phase benchmark reported by \citet{Michaelides2015}, who interpreted thermophoresis in liquids as the ensemble-averaged consequence of temperature-dependent Brownian motion. In the present reproduction of that benchmark, 2000 particles were tracked in a quiescent cubic liquid domain of dimensions \(0.005 \times 0.005 \times 0.005~\mathrm{m}^3\), with an imposed temperature gradient established by fixed wall temperatures. No separate thermophoretic force model was activated. Instead, the ensemble-averaged particle drift velocity was extracted directly from the Lagrangian particle trajectories, such that the observed migration arose solely from the temperature dependence of the Brownian forcing. The comparison shown in figure~\ref{fig:validation_brownian_all}\textit{(b)} demonstrates that the present predictions follow the average trend reported by \citet{Michaelides2015} over the investigated range of temperature gradients and remain within the corresponding confidence interval. This confirms that the Brownian force implementation captures not only ordinary isothermal diffusion, but also the resulting thermophoretic drift in liquid phase systems. This Brownian force model was adopted in all subsequent simulations. Additional information on the implementation within the computational framework is provided Appendix~\ref{appC}. 

\subsection{Particle stress model assessment}
\label{sec:stress_assessment}

The particle stress in the present MPPIC formulation was represented using the Harris-Crighton closure, for which the pressure coefficient \(P_s\) and exponent \(\beta\) were treated as adjustable model parameters. As discussed in section~\ref{sec:formulation}, the appropriate values of these parameters depend on the carrier-phase viscosity and the extent to which inter-particle resistance is represented at the continuum level. For the present nanosuspension problem, suitable values of \(P_s\) and \(\beta\) were determined through an assessment of hindered-settling behaviour at different solid volume fractions.

The benchmark quantity used for this assessment was the normalised mean settling velocity, \(|u_p|/u_{St}\), where \(u_p\) is the average downward particle velocity and \(u_{St}\) is the corresponding single particle Stokes settling velocity. This choice was motivated by the monodisperse hindered-settling results reported by \citet{Li2024}, who considered a non-Brownian, inertialess Stokesian suspension and compared their numerical simulations against the analytical models of \citet{Batchelor1977}, \citet{RICHARDSON1997} and \citet{Hayakawa1995}. Here, the normalised settling velocity was shown to decrease monotonically with increasing solid volume fraction, with Batchelor's model accurate only in the dilute regime ($\phi\ll0.01$). The Hayakawa-Ichiki model, however, provided improved agreement at moderate concentrations ($\phi\approx0.05$). 

\begin{figure}
    \centering
    \includegraphics[width=1.0\textwidth]{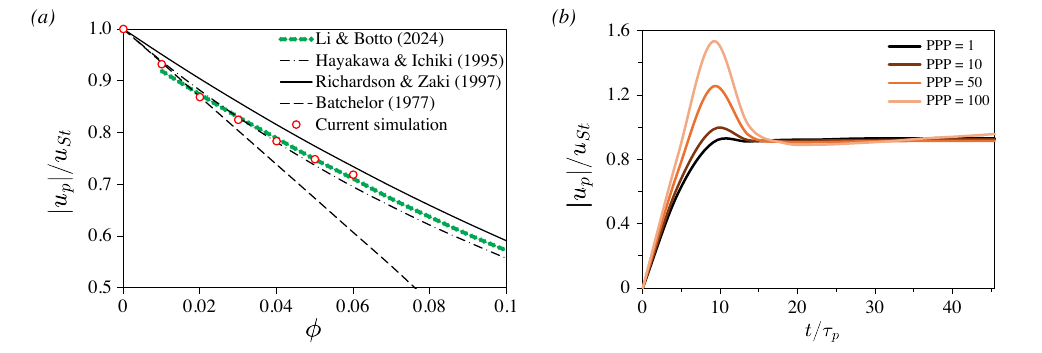}
\caption{Validation of particle stress and particles-per-parcel parameters. (\textit{a}) Assessment of the particle stress model (eq. \ref{eq:tau_p}) on the hindered settling of non-Brownian particles. The present results show the average normalised settling velocity, $|u_p|/u_{St}$, against particle volume fraction, $\phi$, for \(P_s=525\) and \(\beta=1.2\). These are compared with previous analytical models by \cite{Batchelor1977}, \cite{RICHARDSON1997}, \cite{Hayakawa1995}, and the numerical predictions of \citet{Li2024}. (\textit{b}) The sensitivity of particles-per-parcel $(\mathrm{PPP})$ on the predicted temporal evolution of the normalised settling velocity for \(\mathrm{PPP}=1\), 10, 50 and 100.}
\label{fig:hindered_PPP}
\end{figure}

A series of simulations on particle suspension settling were performed over a range of particle volume fractions and a broad sweep of the Harris-Crighton parameters, \(P_s\) and \(\beta\). For each case, the average downward particle velocity was extracted and normalised by \(u_{St}\), and the resulting hindered-settling curve was compared with the reference trend of \citet{Li2024}. Among the combinations tested, \(P_s=525\) and \(\beta=1.2\) produced the best overall agreement. As shown in figure~\ref{fig:hindered_PPP}\text{(a)}, the predictions capture the expected decay in \(|u_p|/u_{St}\) with increasing \(\phi\), and agree with the reference data over the range of concentrations considered. In particular, the resulting curve follows the reference trend closely in the dilute and moderately concentrated regimes, where the hindered-settling behaviour is most relevant to the Taylor-flow suspensions investigated in this work.

On this basis, \(P_s=525\) and \(\beta=1.2\) were adopted in all subsequent simulations. It should be emphasised, however, that this exercise is best interpreted as an assessment and calibration of the particle-stress closure rather than a full validation of the complete solver, since the benchmark of \citet{Li2024} corresponds to idealised non-Brownian Stokesian settling, whereas the present model is intended for fully coupled particle-laden gas-liquid-solid flows.

\subsection{Particles-per-parcel sensitivity}
\label{sec:ppp_sensitivity}

A coarse-graining approach was considered to support the simulation of a considerable number of nanoparticles present in this Taylor flow problem ($\sim10^9$). This approach grouped individual nanoparticles into a numerical parcel. The sensitivity of the MPPIC formulation to the number of physical particles represented by each parcel was assessed in order to determine a suitable particles-per-parcel (PPP) value for the Taylor flow simulations. In this assessment, the baseline solution corresponded to \(\mathrm{PPP}=1\), and progressively coarser parcel representations were then considered.

As a primary indicator, the normalised mean settling velocity, \(|u_p|/u_{St}\), was used to provide a direct measure of the particle dynamics and response to coarsening. The test case considered a particle diameter of \(d_p=100~\mathrm{nm}\), a particle volume fraction of \(1\%\), and an Eulerian cell-size-to-particle-size ratio of 30 (see Appendix \ref{appD} for details on the determination of cell-size-to-particle-size ratio). The numerical parameter of \(\mathrm{PPP}=1\), 10, 50 and 100 were examined, and the temporal evolution of \(|u_p|/u_{St}\) is shown in figure~\ref{fig:hindered_PPP}\textit{(b)}.

The results show that the cases with \(\mathrm{PPP}=1\), 10 and 50 all evolve toward bounded quasi-steady values, although the transient overshoot increases with increasing parcel loading. By contrast, the \(\mathrm{PPP}=100\) case does not reach a settled asymptotic value within the time interval considered. After the initial transient, \(|u_p|/u_{St}\) continues to increase at later times, indicating that this level of coarsening introduces an artificial acceleration into the settling dynamics.  

A secondary assessment was also performed using predictions of the wall-averaged Nusselt number from Taylor flow simulations. This check led to the same qualitative conclusion as the settling-based analysis. Here, \(\mathrm{PPP}\leq 50\) preserved the coupled thermal response with acceptable accuracy relative to the \(\mathrm{PPP}=1\) reference. On this basis, \(\mathrm{PPP}=50\) was taken as the maximum particles-per-parcel value implemented in the numerical model. This choice provided a satisfactory compromise between computational efficiency and numerical reliability, while remaining consistent with both the particle dynamics and heat transfer sensitivity indicators.

\subsection{Solver validation}
\label{sec:validation}

\begin{figure}
    \centering
    \includegraphics[width=1.0\textwidth]{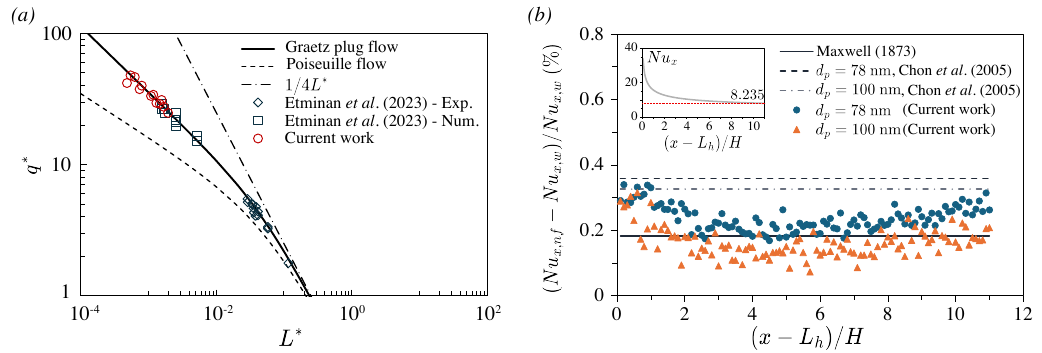}
    \caption{Validation of the current numerical model against previous experimental, analytical, and numerical results. (\textit{a}) Dimensionless mean wall heat flux, $q^*$, versus thermal length, $L^*$, for Taylor flows. Comparisons are made against Graetz flow and Poiseuille flow theory, and the experiments and numerical predictions of \cite{Etminan2023}. (\textit{b}) Deviation of the local Nusselt number, $Nu_x$, for Poiseuille flow comprising water (\textit{w}) and dilute Al$_2$O$_3$--water nanofluid dispersions (\textit{nf}) with $d_p = 78~\mathrm{nm}$, $d_p = 100~\mathrm{nm}$, $\phi = 10^{-4}$, and $Re = 1000$. Comparisons are made against the analytical and empirical models by \cite{Maxwell1873} and \cite{Chon2005}. Inset: Predicted local Nusselt number along the heated channel section showing the thermal development and recovery of the analytical fully developed limit, $Nu_{fd} = 8.235$.}
    \label{fig:validation_all}
\end{figure}

\noindent To verify the accuracy of the Eulerian-Lagrangian framework developed in this work, a validation of the complete numerical method was performed. Two benchmarks were considered to assess the ability to predict heat transport from Taylor flows (absent of nanoparticles) and Poiseuille flows containing nanoparticles (figure \ref{fig:validation_all}). Here, we have focused on non-dimensional heat transfer predictions, which inherently capture the coupled thermal and hydrodynamic characteristics of these flows. In previous work on isothermal Taylor flows, we have shown that bubble topology and thin film thickness are accurately simulated to within $8\%$ of experiments using this approach \citep{Khaleel2025}. The treatment of the particle-interface interactions was also validated through an assessment of the fluidisation of non-Brownian particles in a bubbling, three-phase flow system \citep{Khaleel2025}.    

The dimensionless mean wall heat flux, $q^* = q''_wD_h/(T_w-T_{in})$, for the Taylor flow predictions are presented in figure \ref{fig:validation_all}\textit{a} and compared against previous experiments and simulations by \cite{Etminan2023} and thermal lengths, $L^* = L_{th}/D_hRe Pr$, spanning $L^*\sim10^{-4}-10^{-1}$. Over the range of $L^*$ considered in this work ($L^*\sim10^{-4}-10^{-2}$), the current simulations are in agreement and fall onto the theoretical curve for Graetz plug flow. This is a well-known characteristic of Taylor flows owing to the flow segmentation and internal slug recirculation effects which produce convective heat transport mechanisms that are analogous to plug flow in the small $L^*$ limit ~\citep{Muzychka2010}.    

Our investigations involve the transport of nanofluids from an unheated starting length, $L_h$, through to a heated, thermally developing region (figure \ref{fig:schematic}). We examined the suitability of the numerical method at predicting nanoparticle transport in a similar thermally developing arrangement by considering Poiseuille flow with and without nanoparticles present. At dilute particle concentrations, forced convective heat transport has been found to behave as a single-phase Poiseuille flow with effective thermophysical properties used to describe the nanofluid suspension \citep{Buongiorno2006}. We employ our model to predict the local Nusselt number, $Nu_x$, for thermally developing, Poiseuille flows, and compare it to the solutions obtained using analytical and empirical models for effective properties in dilute conditions (figure \ref{fig:validation_all}\textit{b}). The configuration is a plane channel of gap $H = 1~\mathrm{mm}$ with both walls at uniform heat flux, $Re = 1000$, and $Pr = 7$. $\mathrm{Al_2O_3}$–water dispersions were assessed at $\phi = 10^{-3}$ and particle diameters of $d_p = 78~\mathrm{nm}$ and $d_p = 100~\mathrm{nm}$, respectively.

The analytical model by \cite{Maxwell1873} is independent of particle size (eq. \ref{thermal_conductivity_Maxwell}), predicting equivalent local Nusselt numbers, $Nu_{x,nf}$, for both particle diameters. When compared against the local Nusselt number for the base fluid (water, $Nu_{x,w}$) along the streamwise coordinates of the heated section, $(x-L_h)/H$, an increase in non-dimensional heat transfer of $0.18\%$ is observed (figure \ref{fig:validation_all}\textit{b}). The model developed by \cite{Chon2005} contains particle size dependence, and was obtained using experimental measurements of the thermophysical properties of $\mathrm{Al_2O_3}$–water nanoparticle dispersions for $11~\mathrm{nm}\leq d_p \leq 150~\mathrm{nm}$ and $294~\mathrm{K} \leq T \leq 344~\mathrm{K}$. Here, particle size is shown to influence the non-dimensional heat transfer, with the enhancement in local Nusselt number increasing from $+0.33\%$ to $+0.36\%$ when particle diameter is reduced from $d_p = 100~\mathrm{nm}$ to $d_p = 78~\mathrm{nm}$.

Comparisons were made between these continuum models and the predictions from the current numerical method, which resolves the nanoparticle momentum and energy transport directly. These comparisons, shown in figure \ref{fig:validation_all}\textit{b}, demonstrate an average enhancement to $Nu_x$ along the thermally developing channel section of $+0.17\%$ when $d_p = 100~\mathrm{nm}$, increasing to $+0.24\%$ for the smaller, $d_p = 78~\mathrm{nm}$ particles. Owing to the stochastic nature of the particle transport, the numerical predictions are noisy, yet clearly demonstrate the same sensitivity of $Nu_x$ to particle size effects observed in the empirical model \citep{Chon2005}. The numerical model also demonstrates a sensitivity in $Nu_x$ to the dimensionless thermal coordinate, with the entrance region, $(x-L_h)/H < 1$, and fully developed limit, $(x-L_h)/H > 10$, both leading to local increases in $Nu_x$ enhancement. Considering the numerical method directly resolves Brownian and thermophoretic effects in the flow configuration, this variation is not surprising. They correspond to regions where the thermal boundary layer first emerges, and where it combines downstream to form a fully developed temperature profile. The effect of these thermal adjustments on particle transport phenomena are not considered within either of the analytical or empirical models. Therefore, the current numerical method demonstrates a reasonable agreement with these benchmark models for nanoparticle suspensions and the two-phase Taylor flow phenomena discussed earlier.

\section{Results and discussion}\label{sec:results}

\begin{figure}
    \centering
    \includegraphics[width=1\textwidth]{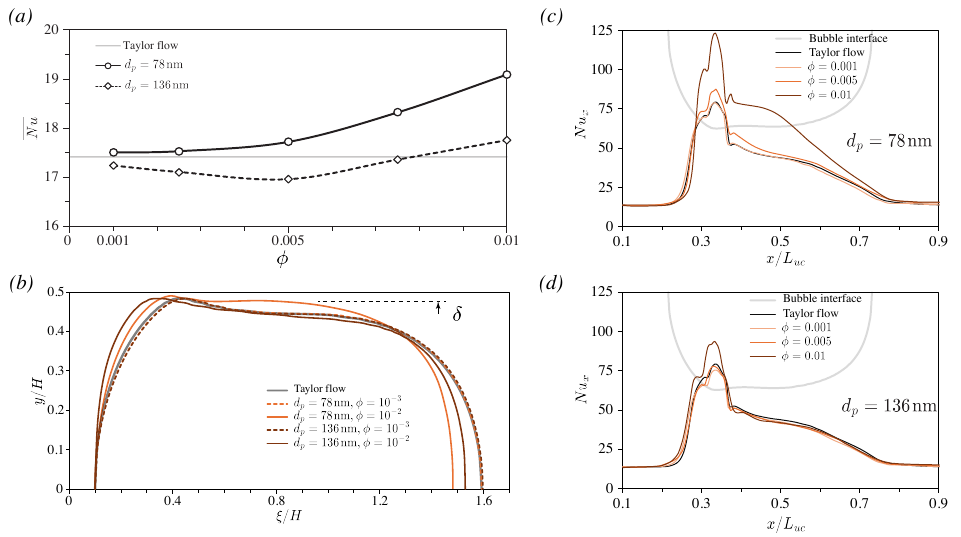}
    \caption{Evolution of dimensionless heat transport and bubble shape with nanofluid particle size and concentration in Taylor flows with $Re = 1000$ and $Ca = 6.9 \times 10^{-3}$. (\textit{a}) Average Nusselt number for a heated section $0.1 \leq L_{th}/H \leq 11$; (\textit{b}) Taylor bubble topology; and (\textit{c,d}) local Nusselt number across the bubble-slug unit cell region, $0.1 \leq L_{uc}/H \leq 0.9$, for $d_p = 78~\mathrm{nm}$ and $d_p = 136~\mathrm{nm}$.}
    \label{fig:Nu_v_dp_phi}
\end{figure}

We begin our discussion by exploring the influence that two key nanofluid parameters have on convective transport in Taylor flow suspensions. Nanoparticle diameter ($d_{p}$) and initial particle concentration ($\phi$) were varied to examine how particle size ($\sim10^1-10^2~\mathrm{nm}$) and concentration ($\sim10^{-3}-10^{-2}$) affect the dimensionless Nusselt number (figure \ref{fig:Nu_v_dp_phi}). The reference conditions considered were $Re = 1000$, $Ca = 6.9 \times10^{-3}$, a wall-normal temperature gradient of $dT/dn = 1 \times 10^{5}~\mathrm{K/m}$, a void fraction of $\varepsilon = 0.5$, and a unit-cell length of $L_{uc}/H = 2.66$, respectively.

We find that the Nusselt number response to increasing particle loading depends strongly on particle size. For $d_p=78~\mathrm{nm}$, the variation is relatively small ($< 1\%$) at the lower concentration limit, followed by a pronounced increase in $\overline{Nu}$ as the concentration approaches $\phi = 10^{-2}$ (figure \ref{fig:Nu_v_dp_phi}\textit{a}). Here, heat transfer enhancement reaches $9.8\%$ relative to the particle-free Taylor flow reference. For the larger $d_p=136~\mathrm{nm}$ particle, however, increasing particle concentration initially reduces $\overline{Nu}$ to $\approx 3\%$ below the Taylor flow reference. The Nusselt number recovers only at the highest concentration, where $\phi=10^{-2}$ produces an enhancement of $2\%$. These results demonstrate that increasing particle loading alone does not guarantee improved heat transfer; its effect depends strongly on the nanoparticle-scale transport mechanisms, and alterations to the Taylor bubble flow at the channel and thin film scales (figure \ref{fig:Nu_v_dp_phi}\textit{b}). 

The local origin of these differences are illustrated in the local Nusselt number distributions shown in figure \ref{fig:Nu_v_dp_phi}\textit{c,d}. For $d_p=78~\mathrm{nm}$, the predominant enhancement at $\phi=10^{-2}$ is concentrated primarily around the bubble end-cap and along the thin-film region. The main peak in $Nu_x$ within the end-cap zone increases by $\approx67\%$, and the enhancement remains appreciable throughout the film region. In contrast, particle concentrations of $\phi\sim10^{-3}$ are much closer to the particle-free Taylor flow response.

A different behaviour is observed for $d_p=136~\mathrm{nm}$ in
figure \ref{fig:Nu_v_dp_phi}\textit{d}. When $\phi\sim10^{-3}$, heat transport from particle-laden flows remain below the Taylor flow reference over most of the film. Increasing the concentration to $\phi =10^{-2}$ produces a notable enhancement to $Nu_x$ that is situated around the bubble end-cap region. Despite this change, the local response remains comparable to, or slightly below, the particle-free flows over much of the remaining thin film. The modest increase in $\overline{Nu}$ at $d_p=136~\mathrm{nm}$ (figure \ref{fig:Nu_v_dp_phi}\textit{a}) is therefore driven exclusively from this localised effect compared to that obtained by the smaller $d_p = 78~\mathrm{nm}$ particles.

Examining the bubble shape with respect to particle diameter and concentration (figure~\ref{fig:Nu_v_dp_phi}\textit{b}), these features are found to be clearly related to three-phase hydrodynamic interactions. At the upper particle concentration threshold of $\phi\sim10^{-2}$, the Taylor bubble topology is altered. Since the bubble lengths differ slightly among the cases, a translated axial coordinate is introduced to allow direct comparison between bubble shapes:
 
\begin{equation}
    \xi/H
    =
    x/H-x_\mathrm{ref}/{H},
    \label{eq:relative_axial_coordinate}
\end{equation}

\noindent where $x_{\mathrm{ref}}$ denotes the selected axial reference position for each bubble. 

\begin{figure}
    \centering
    \includegraphics[width=1\textwidth]{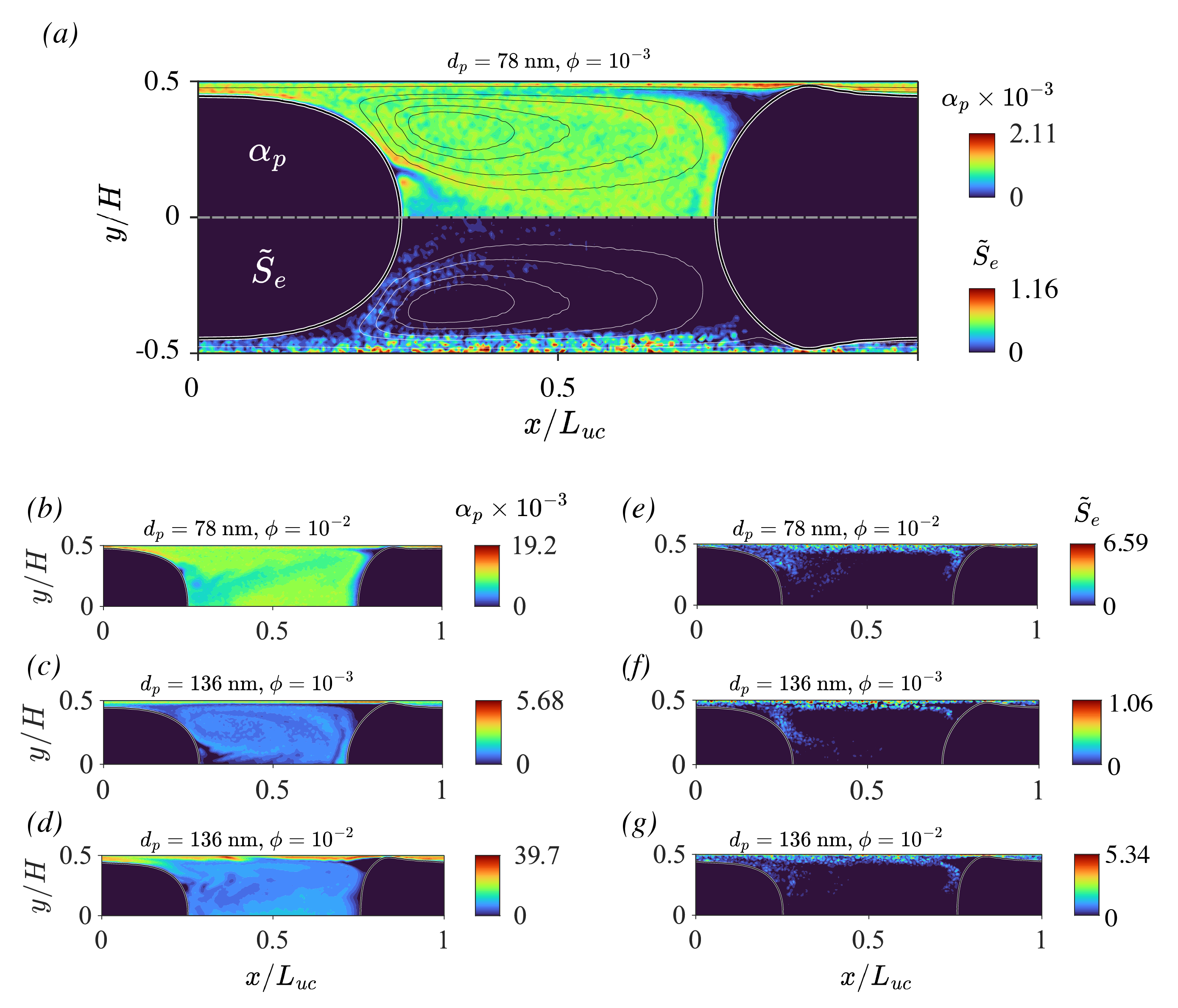}
    \caption{Local particle volume fraction, $\alpha_p$ (\textit{a-d}) and particle-fluid energy exchange, $\tilde{S}_e$ (\textit{a,e-g}) in Taylor flows for $Re = 1000$, $Ca = 6.9 \times 10^{-3}$, and varying particle diameter and initial concentration.}
    \label{fig:phi_Se_dp_phi_intf}
\end{figure}

At $\phi=10^{-3}$, the interface profiles for both particle sizes remain close to that of particle-free Taylor flow. Much larger changes appear at $\phi=10^{-2}$, where the bubble length decreases for both particle sizes. Interestingly, when $d_p=78~\mathrm{nm}$ and $\phi=10^{-2}$, this change in bubble length is accompanied by a $>50\%$ reduction in the liquid film thickness over the central part of the bubble. This provides an additional explanation for the strong local Nusselt number enhancement observed in figure~\ref{fig:Nu_v_dp_phi}\textit{c}, since the reduced wall-to-interface distance decreases the thermal resistance across the film. This hydrodynamic behaviour is also contrary to what would be expected using traditional effective viscosity formulations and one- or two-fluid approaches. Such models relate an increase in $\phi$ with an increase in effective viscosity of the nanofluid \citep{Einstein1905}. Considering for simplicity Bretherton's description of the film thickness, $\delta/H\sim Ca^{2/3}$ \citep{Bretherton1961}, these unresolved modelling approaches imply an increase in film thickness. On the contrary, an increase in effective viscosity can also lower $Re$ and increase the threshold Capillary number for the onset of visco-inertial effects, $Ca^* \sim (\mu^2/\sigma\rho{H})^{3/4}$, where film thickening effects emerge \citep{Quere2000}. However, given the present flows are well within the visco-inertial regime ($Re = 1000$), and the changes in effective thermophysical properties suggested, this does not account for the film thinning observed. Instead, by directly solving for the particle motion in the continuous liquid phase, we observe the importance of the coupled thermal and hydrodynamic interactions within this three-phase system (figure \ref{fig:phi_Se_dp_phi_intf}\textit{a}). 

The dispersion containing larger, $d_p = 136~\mathrm{nm}$ particles exhibits a different topological response: the higher concentration modifies the curvature of the bubble end-cap but does not produce the same pronounced mid-film thinning. This distinction is consistent with the significantly lower heat transfer enhancement obtained for the larger particles ($\approx -80\%$, figure \ref{fig:Nu_v_dp_phi}\textit{a}), and the location of the localised peak in $Nu_x$ observed in figure \ref{fig:Nu_v_dp_phi}\textit{d}. This peak occurs in the end-cap zone where the minimum film thickness is the same as Taylor flows with $\phi \leq 10^{-3}$. However, this minimum occurs over a film length that extends by $\approx 50\%$ further in the stream-wise direction over the Taylor flow reference (figure \ref{fig:Nu_v_dp_phi}\textit{b}).

To elucidate the coupled thermal and hydrodynamic mechanisms underpinning these Nusselt number observations, we investigated the local particle volume fraction, $\alpha_p$, and the magnitude of the interphase energy exchange between nanoparticles and the continuous phases, $\tilde{S}_e$. These distributions are shown in figure \ref{fig:phi_Se_dp_phi_intf} for a single unit-cell (bubble and slug) and different combinations of $d_p$ and initial particle concentration, $\phi$. Comparing equivalent particle volume fractions (figure \ref{fig:phi_Se_dp_phi_intf}\textit{a} and \ref{fig:phi_Se_dp_phi_intf}\text{c}), there is a notable change in particle homogeneity within the slug and near-wall regions when $d_p<100~\mathrm{nm}$ or $d_p>100~\mathrm{nm}$. For $d_p=78~\mathrm{nm}$, Brownian transport maintains a comparatively well-distributed particle population throughout the recirculating liquid slug (figure \ref{fig:phi_Se_dp_phi_intf}\textit{a,b}). Increasing $\phi$ does not strengthen the Brownian forcing acting on an individual particle; rather, it increases the number of particles participating in Brownian redistribution and particle-fluid thermal exchange. Consequently, the high concentration case (figure \ref{fig:phi_Se_dp_phi_intf}\textit{b}) contains a larger population of particles that can be repeatedly transported between the slug interior, bubble vicinity, and near-wall region by the recirculating slug flow.

The consequence of an increased particle loading on particle-fluid energy exchange is reflected in the corresponding $\tilde{S}_e$ fields shown in figure \ref{fig:phi_Se_dp_phi_intf}\textit{a,e} for $d_p = 78~\mathrm{nm}$. At $\phi_p=10^{-2}$, the local particle-fluid energy exchange increases substantially ($\approx5.7\times$) and is distributed throughout the recirculating liquid slug region, approaching the channel centreline at $y/H = 0$. Increasing particle concentration also introduces a thermophysical trade-off: the lower specific heat capacity of $\mathrm{Al_2O_3}$ reduces the effective heat-storage capacity of the suspension, whereas the higher particle thermal conductivity coupled with the Brownian transport mechanism increases the effective thermal conductivity of the nanofluid \citep{Chon2005}. The heat transfer response therefore depends on the balance between these competing effects and on how effectively the particles are transported within the liquid slug and thin film flows. For $d_p=78~\mathrm{nm}$, the stronger Brownian redistribution, together with slug recirculation and a larger particle population, promotes more widespread particle--fluid thermal exchanges and allows this effective conductivity benefit to become more pronounced. Thus, the concentration dependence of the Nusselt number reflects the combined influences of bubble topology changes, particle redistribution forces from nano- (Brownian) to channel length-scales (convective Taylor flow circulations), and the associated solid-liquid interphase energy exchanges.

For the dispersions containing larger $d_p=136~\mathrm{nm}$ particles, Brownian diffusion is substantially weaker. Increasing the particle loading does not produce the same degree of redistribution through the slug as for the smaller particles. The concentration fields retain greater spatial non-uniformity, while appreciable particle populations remain in the near-wall and thin film regions (figure \ref{fig:phi_Se_dp_phi_intf}\textit{c,d}). Nevertheless, increasing $\phi$ increases the absolute number of particles available for thermal exchange, and the corresponding $\tilde{S}_e$ field strengthens by $\approx5\times$ when the concentration is increased to $\phi=10^{-2}$ (figure \ref{fig:phi_Se_dp_phi_intf}\textit{f,g}). This increased particle-mediated energy exchange, coupled with the alteration of the bubble topology (figure \ref{fig:Nu_v_dp_phi}\textit{b}), is consistent with the recovery of $\overline{Nu}$ at the highest concentration (figure \ref{fig:Nu_v_dp_phi}\textit{a}). A weaker Brownian redistribution and less favourable film thinning behaviour around the Taylor bubble prevents the much larger enhancements to $Nu_x$ and $\overline{Nu}$ observed when $d_p<100~\mathrm{nm}$.

Overall, we observe that the effects of particle concentration are strongly coupled to particle size in Taylor flows. Despite the modest initial concentrations ($\phi \leq0.01$), and local particle volume fractions predicted ($\alpha_p<0.04$), these results also highlight the limitations of classical Eulerian models that are traditionally used to describe the thermophysical properties of nanofluids and non-dimensional heat transfer. Furthermore, when $\phi\gg10^{-3}$, Taylor bubble morphology and liquid film thickness were found to be significantly altered by the particle's presence and size. This sensitive coupling between the three phases poses a challenge to isolate the effect of nanoparticle transport and Taylor flow parameters on non-dimensional heat transport, including Brownian diffusion, thermophoretic drift, particle-interface interactions, Reynolds number, and gas bubble and liquid slug lengths. To address this, we have focused the remainder of the discussions using the low concentration limit ($\phi=10^{-3}$), where Taylor bubble shape and liquid slug circulation are consistent with Taylor flows without nanoparticles (figure \ref{fig:Nu_v_dp_phi}\textit{b}). This will be supported by investigations on Poiseuille flows containing nanofluids, to assess how nanoparticle transport phenomena in the absence of Taylor bubbles alters non-dimensional heat transport, and also provide an insight on the transport behaviour in the long slug limit, $\tilde{L}_s\gg1$ \citep{Leung2010}.

\subsection{Reference flow states and transport mechanisms}
\label{sec:reference_states}

Nanoparticle transport mechanisms were first examined by comparing four reference configurations: Poiseuille flow (PF), particle-laden Poiseuille nanofluid flow (PNF), Taylor flow (TF), and particle-laden Taylor nanofluid flow (TNF). The reference particle-containing cases used an $\mathrm{Al_2O_3}$ particle diameter of $d_p=78~\mathrm{nm}$ and an initial particle volume fraction of $\phi=10^{-3}$. All other parameters were maintained as before, including the wall-normal temperature gradient $(dT/dn=1\times10^{5}~\mathrm{K/m})$, Reynolds number $(Re=1000)$, Capillary number $(Ca = 6.9 \times 10^{-3})$, void fraction $(\varepsilon=0.5)$, and the unit-cell length $(L_{uc}/H=2.66)$. All instantaneous fields were extracted at the same reference flow state within the heated section, $L_{th}$ (figure \ref{fig:schematic}).

\begin{figure}
    \centering
    \includegraphics[width=1\textwidth]{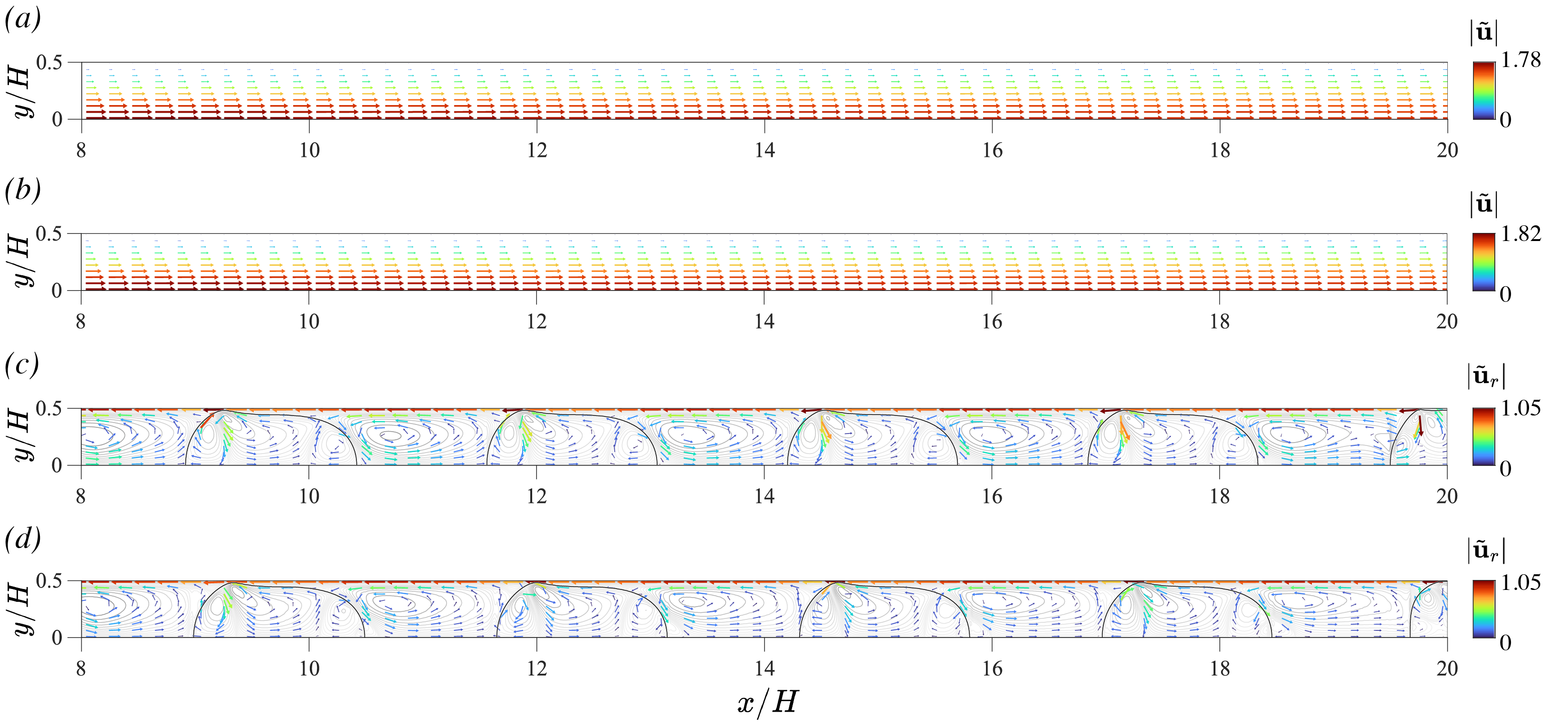}
    \caption{Instantaneous velocity fields within the heated section, $L_{th}$, for the four reference configurations: (\textit{a}) Poiseuille flow, (\textit{b}) Poiseuille nanofluid flow, (\textit{c}) Taylor flow, and (\textit{d}) Taylor nanofluid flow. The Taylor flows are presented in the translating bubble reference frame to reveal the internal bubble and slug circulations, where $|\tilde{\mathbf{u}}_r| = \rho |\mathbf{u}_r|D_h / \mu Re$, and $\mathbf{u}_r = \mathbf{u}-\mathbf{u}_b$.}
    \label{fig:Streamlines & Velocity vectors (section 5.1)}
\end{figure}

Figure~\ref{fig:Streamlines & Velocity vectors (section 5.1)} compares the dimensionless velocity fields of the four configurations, where $|\tilde{\mathbf{u}}| = \rho |\mathbf{u}|D_h / \mu Re$. The Poiseuille flows exhibit the expected unidirectional velocity profiles, with no appreciable cross-stream motion. A marginally higher velocity magnitude is observed in the nanofluid case (figure \ref{fig:Streamlines & Velocity vectors (section 5.1)}\textit{b}). This difference may reflect the small change in the effective mixture properties; however, the similarity of the two velocity fields indicates that the dilute particles do not substantially modify the macroscale hydrodynamics. A fundamentally different flow topology is observed in the Taylor flow cases (figure \ref{fig:Streamlines & Velocity vectors (section 5.1)}\textit{c,d}). Streamlines, evaluated in the translating reference frame, reveal closed recirculation cells within the liquid slugs, together with strong velocity variations near the bubble caps and the liquid films. The convective circulation patterns are nearly identical with and without nanoparticles, indicating that the addition of particles at $\phi =10^{-3}$ has little influence on the resolved gas-liquid flow structure. Consequently, differences between the thermal responses of the two Taylor flow cases are primarily associated with particle transport and particle-fluid thermal and hydrodynamic coupling rather than a substantial modification of the bubble morphology and slug circulation.

\begin{figure}
    \centering
    \includegraphics[width=1\textwidth]{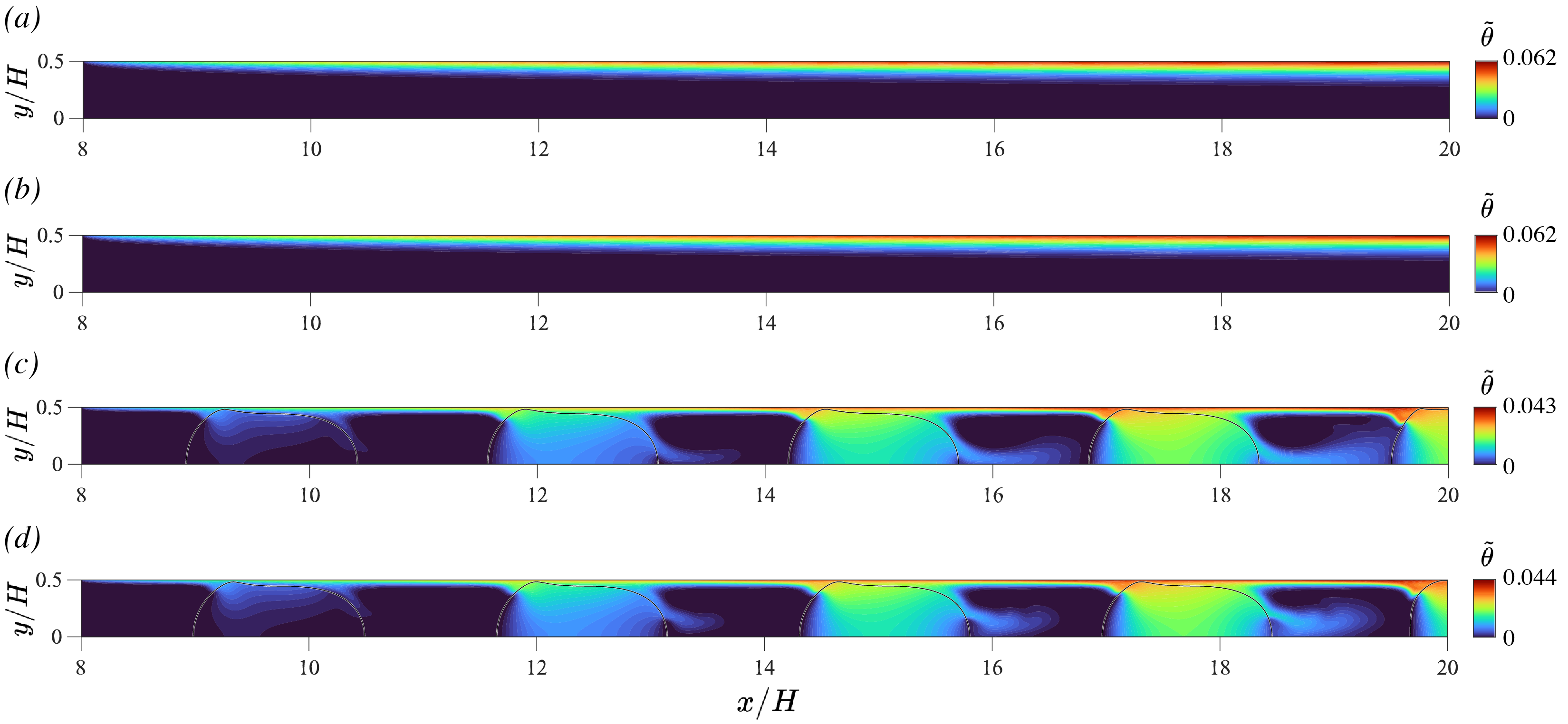}
    \caption{Instantaneous temperature fields within the heated section for (\textit{a}) Poiseuille flow, (\textit{b}) Poiseuille nanofluid flow, (\textit{c}) Taylor flow, and (\textit{d}) Taylor nanofluid flow. The Poiseuille flows show progressive thermal boundary layer development driven by diffusion, whereas the Taylor flows produce periodic displacement of the isotherms by the convective internal bubble-slug circulations.}
    \label{fig:istherms (section 5.1)}
\end{figure}

The corresponding dimensionless temperature fields, $\tilde{\theta} = (T-T_{in})\kappa/q_w''D_h$, are presented in figure~\ref{fig:istherms (section 5.1)}. The thermal boundary layer due to Poiseuille flow develops progressively from the heated wall toward the channel centreline. Only a small difference is visible between the particle-free and particle-laden temperature fields, consistent with their similar velocity distributions. The temperature fields for the Taylor flows are periodically restructured by the bubble-slug sequence. The isotherms are displaced from the wall and transported into the interior of the liquid slugs, following the recirculating motion identified in figure~\ref{fig:Streamlines & Velocity vectors (section 5.1)}. This process continuously renews the near-wall fluid and promotes radial heat transport compared with the predominantly diffusive cross-stream transport of the Poiseuille flows. The addition of nanoparticles produces a modest further modification of the temperature distribution within the liquid slugs. Since the underlying recirculation pattern remains nearly unchanged, this difference is attributed to the redistribution of the particles and their thermo-fluidic interaction with the carrier liquid that will be examined in the following sections (Sec. \ref{sec:Brownian&thermophoretic}-\ref{sec:flow_structure_effects}). The effect is most visible within the recirculating slug regions, where particles can repeatedly exchange energy with the surrounding liquid across the channel height ($0<|y/H|<0.5$). 

\begin{figure}
    \centering
    \includegraphics[width=1\textwidth]{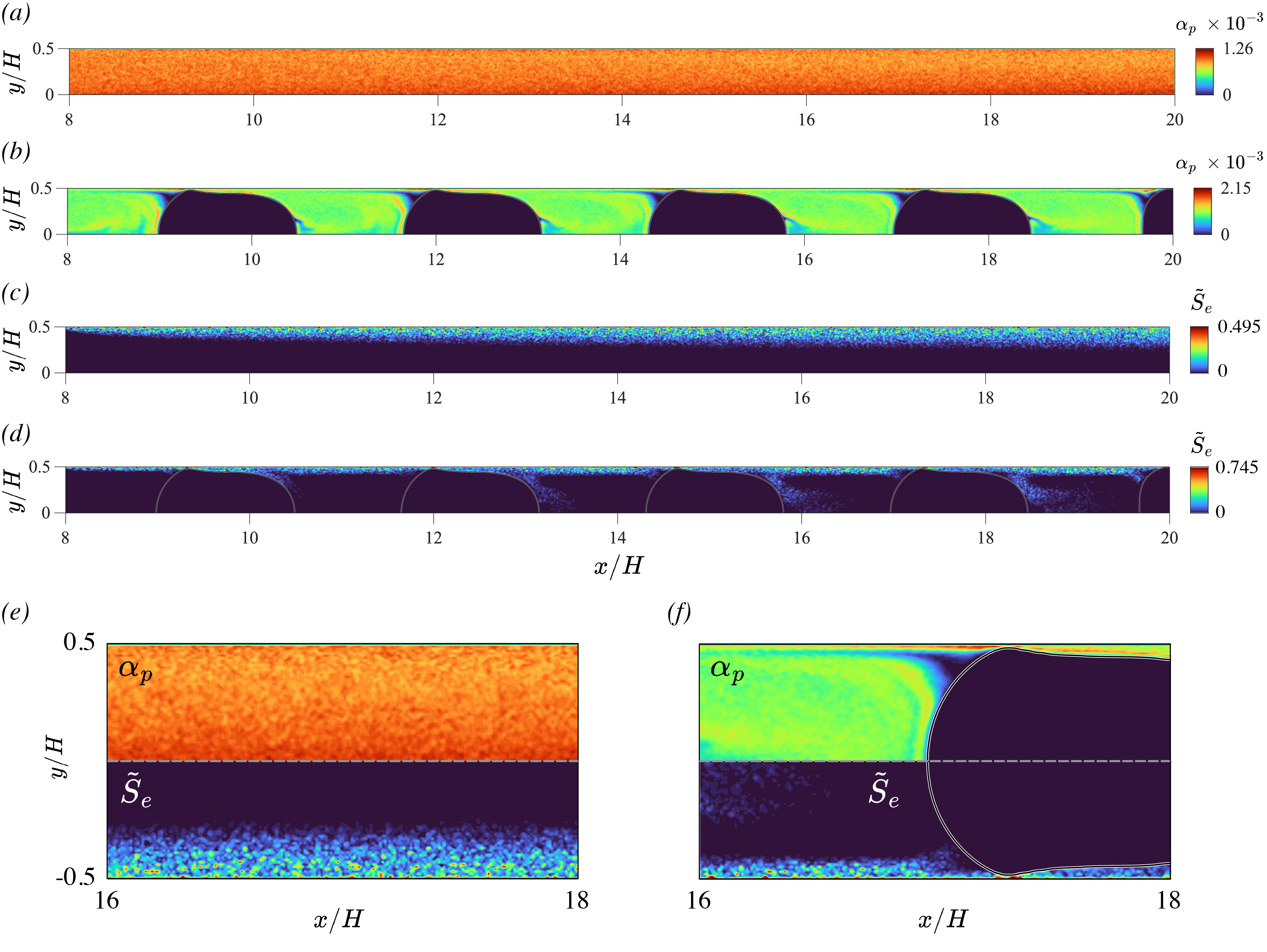}
    \caption{Particle distribution and interphase energy exchanges within the heated section. Local particle volume fraction, $\alpha_p$, and particle-fluid energy exchange, $\tilde{S}_e$, for (\textit{a,c}) Poiseuille flow and (\textit{b,d}) Taylor flow. (\textit{e,f}) Close-up view of a downstream channel section for both flow configurations.}
\label{fig:cosentration&Se (Section 5.1)}
\end{figure}

Figure~\ref{fig:cosentration&Se (Section 5.1)} shows the local particle volume fraction, $\alpha_p$, and the magnitude of the Eulerian interphase energy source, $\tilde{S}_e$, for the two particle-containing flow configurations. This allows us to visualise the particle distributions on the scale of the channel height and heated length ($H,~L_{th}$). In Poiseuille flow, the particle distribution remains approximately homogeneous at the channel scale, apart from the local fluctuations inherent to the Lagrangian representation (figure \ref{fig:cosentration&Se (Section 5.1)}\textit{a,e}). By
contrast, Taylor flows exhibit a strongly non-uniform distribution (figure \ref{fig:cosentration&Se (Section 5.1)}\textit{b,f}). Particle enrichment develops along the near-wall liquid films and around parts of the bubble caps, while the particle distribution within the liquid slugs is shaped by the recirculating flow.

In Poiseuille nanofluid flow, local maxima in $\tilde{S}_e$ are concentrated mainly within the developing near-wall thermal layer (figure \ref{fig:cosentration&Se (Section 5.1)}\textit{c,e}). This is the region in which the particle and carrier-fluid temperatures undergo their strongest local adjustment ($0.4 < |y/H| < 0.5$). In particle-laden Taylor nanofluid flow, regions of appreciable $\tilde{S}_e$ extend into the channel centre ($y/H = 0$) and follow the circulating convective flow structures within the liquid slugs. Taylor circulations therefore redistribute the locations at which nanoparticle-fluid thermal exchange occur, transferring the influence of the heated near-wall region into the slug interior. 

To quantify the relative changes between any investigated configuration, $A$, and a selected reference configuration, $B$, a
non-dimensional heat transfer enhancement parameter has been defined as:

\begin{equation}
\eta_{Nu}(A\mid B)
=
\frac{\overline{Nu}_{A}-\overline{Nu}_{B}}
{\overline{Nu}_{B}},
\label{eq:Nu_enhancement}
\end{equation}

\noindent where $\overline{Nu}$ is given by equation~\ref{eq:Nu_bar}. The four flow configurations have been compared in table~\ref{tab:reference_nusselt} using equation~\eqref{eq:Nu_enhancement} to calculate their relative enhancement values. We find that the wall-averaged Nusselt number increases from $10.253$ to $10.281$ following nanoparticle addition in Poiseuille flows. This corresponds to an enhancement of $0.27\%$, demonstrating that the nanoparticles produce only a small thermal effect in Poiseuille flow at the selected dilute concentration of $\phi=10^{-3}$. This magnitude of enhancement is expected, given the outcomes from our validations in Sec. \ref{sec:implementation}, and previous investigations on nanofluid flow through microfluidic channels at $Re<100$ \citep{Abu-Nada2019}. 

A markedly different response is obtained when bubble segmentation is introduced. Taylor bubbles are shown to generate peaks in interphase energy exchange that are almost two-fold above that observed for Poiseuille flows. These occur around the regions of minimum liquid film thickness, and are translated along the wall length as the bubble is transported through the heated section (figure \ref{fig:cosentration&Se (Section 5.1)}\textit{d,f}). This leads to local peaks in Nusselt number that are over three-fold above the slug region (figure \ref{fig:Nu_v_dp_phi}\textit{c,d}). Heat gained by the thin film and near-wall regions of the slug is then transported effectively into the bulk liquid through the channel-scale convective motion within the liquid slugs (figure \ref{fig:phi_Se_dp_phi_intf}\textit{a}). In the absence of nanoparticles, the wall-averaged Nusselt number increases from $10.253$ for Poiseuille flow to $17.394$ for Taylor flow, representing an enhancement of $69.64\%$ (table~\ref{tab:reference_nusselt}). 

Particle addition produces smaller but noticeable differences between the two Taylor flows (TF, TNF), particularly near the bubble-associated local Nusselt number peaks and thin film regions (figure \ref{fig:Nu_v_dp_phi}\textit{c,d}). These local differences in Nusselt number emerge due to nanoparticle enrichment within the thin film and near-wall regions, evidenced by the distribution of the local volume fraction, $\alpha_p$ (figure \ref{fig:cosentration&Se (Section 5.1)}\textit{f}). This is both driven by, and ultimately intensifies, the particle-fluid thermal exchanges, $\tilde{S}_e$, in the near-wall region of the channel. The large scale convective motion within the slug transports this heated particle-fluid mix across the channel height (figure \ref{fig:phi_Se_dp_phi_intf}\textit{a}). This promotes an additional particle-induced enhancement to the wall-averaged Nusselt number of $0.64\%$ relative to particle-free Taylor flow. Although a modest overall increase, it is $2.37\times$ more than the enhancement observed for Poiseuille nanofluid flows. This suggests that coupled thermal and hydrodynamic interactions in the three-phase Taylor flow configuration play an important role. To understand the origins of these co-benefits, the following sections will examine the influence of Brownian effects, particle-interface interaction, Reynolds number, and slug and bubble parameters in Taylor nanofluid flows.

\begin{table}
\begin{center}
\begin{tabular}{lcccccc}
Case & $Re$ & $Ca$ & $L_{th}/H$ & $\overline{Nu}$ & Reference case & $\eta_{Nu}$ \\[3pt]
Poiseuille flow, PF & 1000 & --- & 12 & 10.253 & --- & --- \\
Poiseuille nanofluid flow, PNF & 1000 & --- & 12 & 10.281 & PF & $0.27\%$ \\
Taylor flow, TF & 1000 & $6.9 \times 10^{-3}$ & 12 & 17.394 & PF & $69.64\%$ \\
Taylor nanofluid flow, TNF & 1000 & $6.9 \times 10^{-3}$ & 12 & 17.506 & TF & $0.64\%$ \\
\end{tabular}
\caption{Wall-averaged Nusselt numbers for the four reference configurations. 
}
\label{tab:reference_nusselt}
\end{center}
\end{table}

\subsection{Brownian diffusion and thermophoretic drift}
\label{sec:Brownian&thermophoretic}

In the present liquid-phase formulation (eq. \ref{eq:Brownian_force}), thermophoresis is not introduced as a separate deterministic force. Instead, it emerges from the spatial dependence of the stochastic Brownian forcing under a temperature gradient (figure \ref{fig:validation_brownian_all}). Here, we have isolated the combined effect of Brownian diffusion and the resulting thermophoretic drift on nanofluid transport by performing simulations with the Brownian model enabled and disabled. To establish the persistent wall-directed migration driven by thermopheretic drift, the time-averaged radial concentration bias was quantified using: 
\begin{equation}
I_{th}(x)
=
\frac{\overline{\alpha}_{p,c}(x)-\overline{\alpha}_{p,w}(x)}
{\phi},
\label{eq:Ith}
\end{equation}
where $\overline{\alpha}_{p,c}$ and $\overline{\alpha}_{p,w}$ represent the particle concentrations averaged over selected centreline ($|y/H| = 0$) and near-wall bands ($|y/H| = 0.49$), respectively. With this definition, positive values of the concentration bias index, $I_{th}$, indicate particle enrichment toward the channel centre relative to the heated, near-wall region.

\begin{figure}
  \centering
  \includegraphics[width=1.0\textwidth]{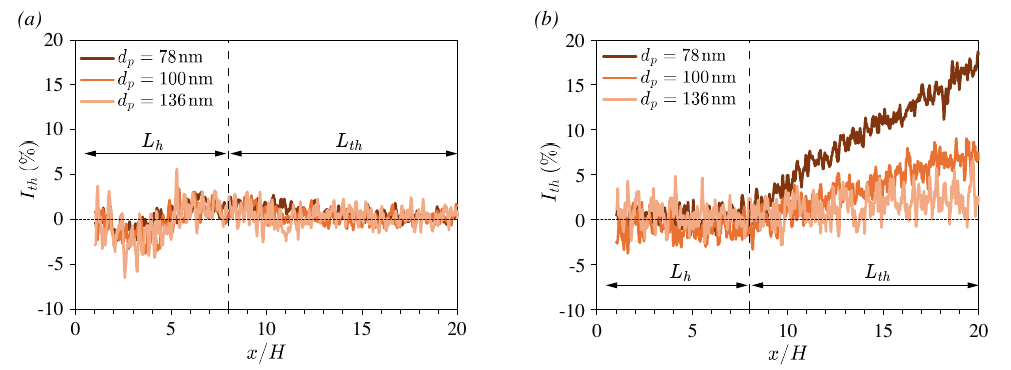}
  \caption{Streamwise evolution of the time-averaged thermophoretic concentration-bias index, $I_{th}$, for $d_p=78$, $100$, and $136~\mathrm{nm}$ particles in Poiseuille flows (\textit{a}) without and (\textit{b}) with Brownian forcing on the particle phase. Positive values indicate centreline enrichment relative to the heated near-wall region.}
  \label{fig:Ith}
\end{figure}

When Brownian forcing is disabled, $I_{th}$ fluctuates around zero after the inlet and hydrodynamic development region ($L_h$), without any systematic dependence on particle diameter (figure \ref{fig:Ith}\textit{a}). Advection alone, therefore, does not produce a persistent wall-to-centre concentration bias in Poiseuille flows. Enabling Brownian forcing through the non-isothermal Brownian model (eq. \ref{eq:Brownian_force}) establishes a different response for the particle transport (figure \ref{fig:Ith}\textit{b}). Beyond $x/H\approx8$, where the heated channel section begins, $I_{th}$ becomes positive and increases downstream, demonstrating particle migration away from the heated wall along the entire heated length, $L_{th}$. This thermophoretic drift strengthens as the particle diameter decreases, where the outlet value approaches $I_{th}\approx 18\%$ for $d_p=78~\mathrm{nm}$, compared with $I_{th}\approx 7\%$ for $d_p=100~\mathrm{nm}$, and only a few percent for $d_p=136~\mathrm{nm}$.

\begin{figure}
  \centering
  \includegraphics[width=1\textwidth]{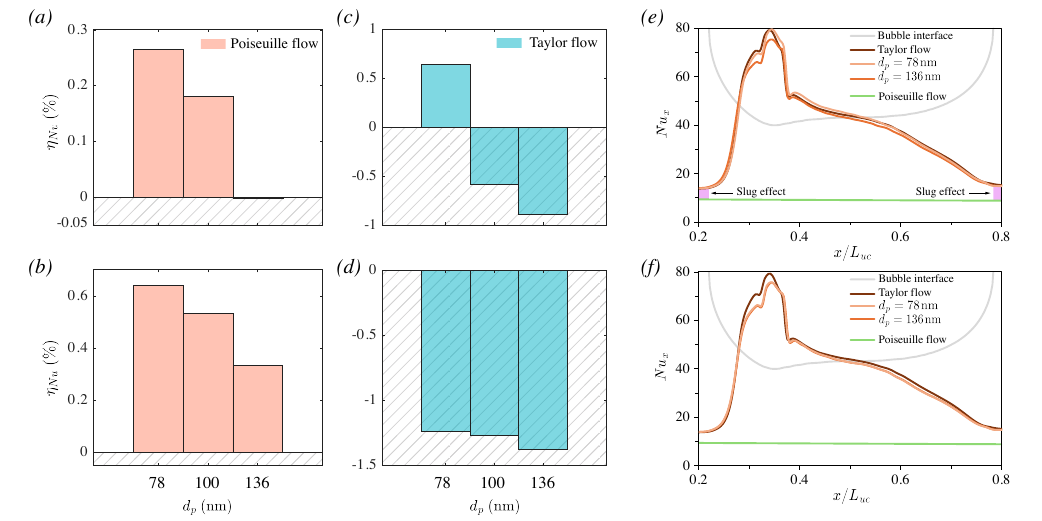}
  \caption{The influence of Brownian effects on average and local Nusselt numbers. Enhancement to wall-averaged Nusselt number for (\textit{a,b}) Poiseuille flow and (\textit{c,d}) Taylor flow. (\textit{e,f}) Local Nusselt number for Taylor flows, also showing Poiseuille flow. The top row are results for Brownian particles (\textit{a,c,e}), while the bottom row are for non-Brownian particles (\textit{b,d,f}).}
  \label{fig:Average Nu (section2)}
\end{figure}

We examined how Brownian motion and thermophoretic drift influence the wall-averaged Nusselt number by comparing against the base flow configurations without nanoparticles ($\eta_{Nu}$, figure \ref{fig:Average Nu (section2)}\textit{a-d}). For Poiseuille flow in the presence of Brownian particles, all nanofluid cases remain close to the water reference at $\phi=10^{-3}$, transitioning from a slight reduction to $\overline{Nu}$ when $d_p = 136~\mathrm{nm}$, to an enhancement as particle diameter is reduced to $d_p\leq 100~\mathrm{nm}$. The non-Brownian predictions show higher $\eta_{Nu}$ than the Brownian-enabled results. This difference is the outcome of the thermophoretic migration which forces particles to move from the higher temperature near-wall region to the cooler bulk fluid toward the channel centreline (figure~\ref{fig:Ith}). When Brownian forcing is omitted, particles remain in the near-wall region, supporting heat transport across the boundary layer. In contrast, the full model, which captures Brownian effects, predicts particle depletion from the boundary layer (figure \ref{fig:Ith}\textit{b}). Neglecting Brownian motion, and the accompanying thermophoretic transport, therefore, over-predicts the nanoparticle enhancement to the Nusselt number in Poiseuille flows.

\begin{figure}
  \centering
  \includegraphics[width=\textwidth]{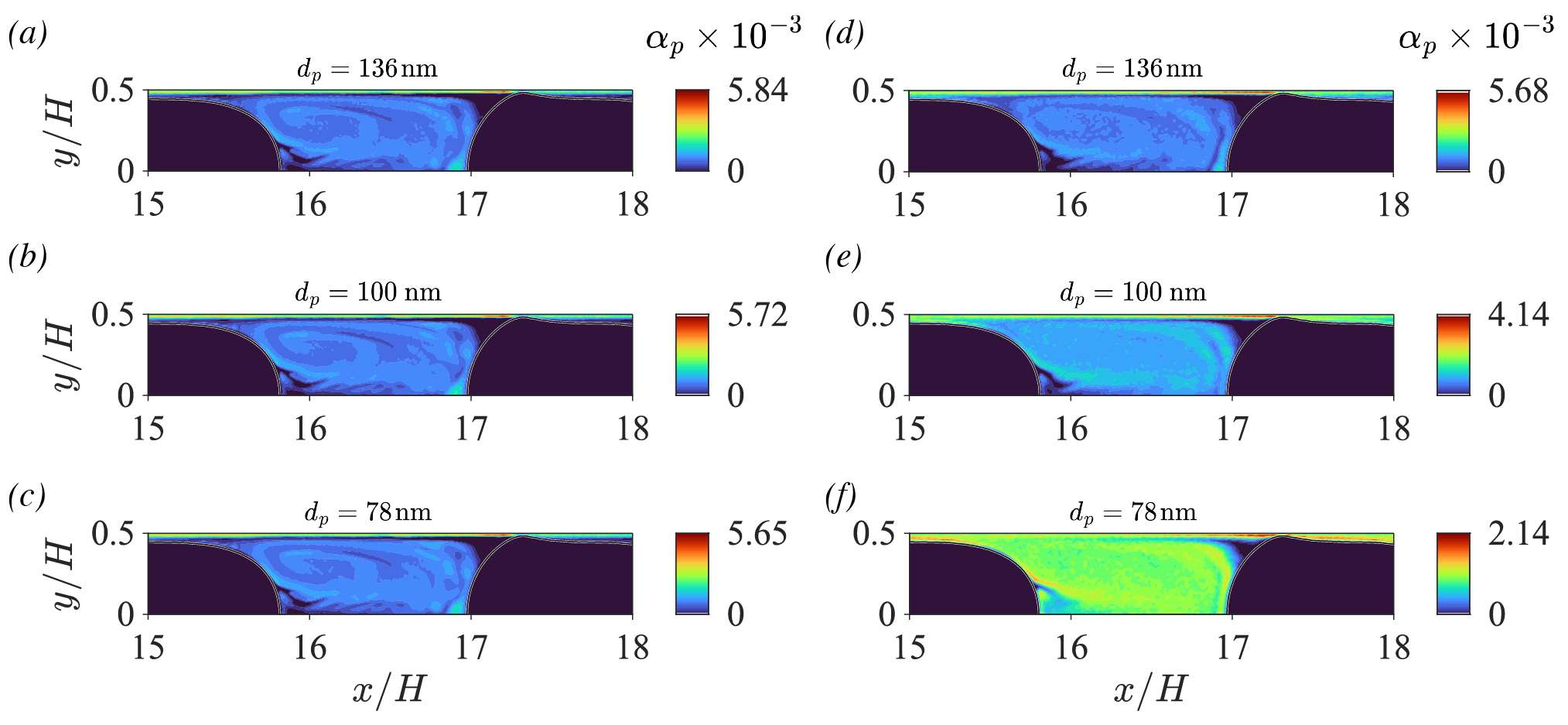}
  \caption{Local particle volume fraction within a representative Taylor flow unit-cell for $d_p=136$, $100$, and $78~\mathrm{nm}$ with
  (\textit{a-c}) Brownian forcing disabled and (\textit{d-f}) Brownian forcing enabled.}
  \label{fig:concentrations (sec. 5.2)}
\end{figure}

Heat transport behaviour in Taylor flow is both qualitatively and quantitatively different. Without Brownian forcing, the wall-averaged Nusselt number in particle-laden Taylor flow depends only weakly on particle diameter and remains below the particle-free Taylor flow reference (figure \ref{fig:Average Nu (section2)}\textit{d}). Spatial distributions of the local particle volume fraction are similar for all sizes (figure \ref{fig:concentrations (sec. 5.2)}\textit{a-c}). Particles enter the thin liquid film around the Taylor bubble and remain in the near-wall region of the slug, establishing a narrow band with increased particle concentration at $|y/H| \approx 0.5 - \delta$. A stagnation streamline exists from the bubble end-cap at $x/H \approx 17$, to the bubble nose at $x/H\approx 15.5$ (figure \ref{fig:Streamlines & Velocity vectors (section 5.1)}\textit{d}). This creates a barrier in particle transport between the concentrated near-wall region and the convective circulations within the slug. This effect is observed by a noticeable $\alpha_p\approx0$ dead-zone which emerges at the bubble rear, and extends with the flow to the front of the following bubble (figure \ref{fig:concentrations (sec. 5.2)}\textit{a-c}). This increase in effective fluid thermal conductivity in the near-wall region, combined with the barrier for particle momentum and energy exchanges with the convective motion inside the slug, leads to the reduction in local and wall-averaged Nusselt number for Taylor flow when Brownian forcing is absent (figure \ref{fig:Average Nu (section2)}\textit{d,f}). 

\begin{figure}
  \centering
  \includegraphics[width=1\textwidth]{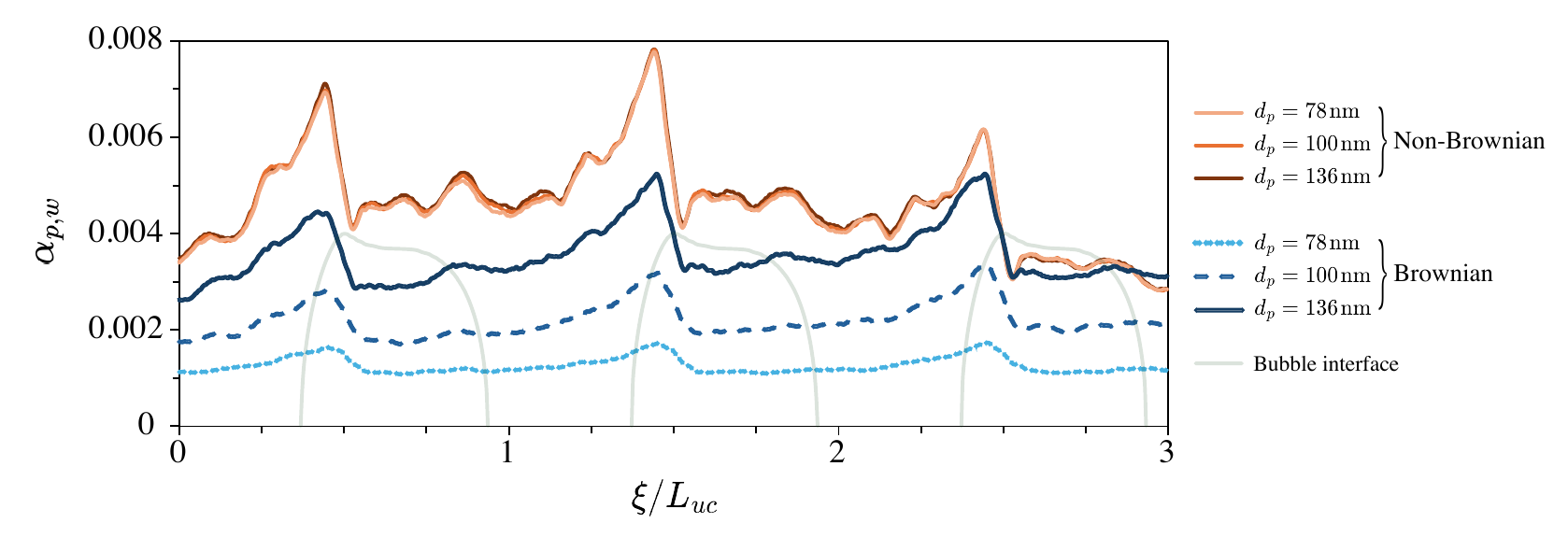}
  \caption{Near-wall particle volume fraction profiles through three successive bubble and film regions for $\phi = 10^{-3}$, and $d_p=78$, $100$, and $136~\mathrm{nm}$, with Brownian forcing enabled and disabled.}
  \label{fig:film_concentration}
\end{figure}

With the Brownian model enabled, a pronounced size dependence appears for the local particle volume fraction fields (figure \ref{fig:concentrations (sec. 5.2)}\textit{d-f}) and this alters the non-dimensional heat transport (figure \ref{fig:Average Nu (section2)}\textit{c,e}). The largest particle has a similar $\alpha_p$ distribution to the non-Brownian counterpart (figure \ref{fig:concentrations (sec. 5.2)}\textit{a,d}), with a $\approx20-30\%$ reduction in the amount of particles concentrated in the near-wall region (figure \ref{fig:film_concentration}), and improved particle exchanges across the stagnation streamline separating this near-wall region with the flow circulations. Particle redistribution from the film region to the slug continues to improve as particle size is reduced further (figure \ref{fig:concentrations (sec. 5.2)}\textit{e,f}, figure \ref{fig:film_concentration}). Brownian motion and thermophoretic drift effects increase with decreasing particle diameter, allowing particles to transport across streamlines, reducing the size of the $\alpha_p\approx0$ zone at the bubble rear ($x/H\approx 17$). Using the smallest particle diameter studied in this work, $d_p=78~\mathrm{nm}$, the highest average Nusselt number is obtained, exceeding the baseline particle-free Taylor flow (figure \ref{fig:Average Nu (section2)}\textit{c}). Here, convective motion within the liquid slug continually redistributes particles across the channel (figure \ref{fig:phi_Se_dp_phi_intf}\textit{a}), opposing the persistent wall-to-centre thermophoretic particle segregation observed in Poiseuille flows. 

The local Nusselt number profiles in figure~\ref{fig:Average Nu (section2)}\textit{e,f} provide a spatial interpretation of the averaged response. In the non-Brownian cases, the curves for the $d_p=78$ and $136~\mathrm{nm}$ particles nearly collapse and remain below the particle-free Taylor flow profile over most of the bubble region. Enabling Brownian forcing separates the two particle cases: the $78~\mathrm{nm}$ profile is higher than the $136~\mathrm{nm}$ profile over a substantial part of the liquid film region, and locally matches or exceeds the particle-free Taylor flow response. The diameter dependence of the average Nusselt number therefore arises from localized changes around the bubble film and rear-meniscus region rather than from a uniform shift along the complete unit cell. This is reflected by $Nu_x$ in the slug regions, near $x/L_{uc} = 0.2,~0.8$, where changes are relatively minor (figure \ref{fig:Average Nu (section2)}\textit{e}). These findings indicate that, unlike Poiseuille flows, it is the Brownian effects coupled with the unique hydrodynamics of Taylor flows, which produce Nusselt number enhancements for nanofluid suspensions.

\subsection{Particle--interface interaction effects}
\label{sec:interface_coefficient}

To examine if changes in nanoparticle interactions with a Taylor bubble can alter hydrodynamic and heat transport phenomena, we performed an assessment on the particle-interface force coefficient, $C_{int}$. This coefficient controls the strength of particle interaction with the gas-liquid interface (eq. \ref{eq:Fint_alphaH}). Although this force model is a simplification to prevent particle attachment at the interface, and describe a net repulsive force due to several underlying mechanisms (van der Waals, electric double-layer, Hydration, eq. \ref{eq:net_int_F}), it allows us to investigate potential changes in repulsion strength that could arise when sensitive nanofluid dispersion parameters are modified (e.g. pH around the isoelectric point, Appendix \ref{appB}). Here, the coefficient was varied over $5 \leq C_{int}\leq 25$ for particle diameters of $100~\mathrm{nm}$ and $136~\mathrm{nm}$. Values below $C_{int}=5$ allowed particles to penetrate into the gas phase for the smallest $d_p = 78~\mathrm{nm}$ particle, and were therefore excluded from this investigation on the particle-interface interactions. This resulted in a range of interface forces spanning $0.38~\mathrm{pN} \leq\mathbf{F}_{int}\leq 7.5~\mathrm{pN}$, which reflects physically realistic conditions for this $\mathrm{Al_2O_3}$-water nanofluid (Appendix \ref{appB}). As discussed in Sec. \ref{sec:Interface-force correction}, it represents the net short-range repulsion between a hydrophilic nanoparticle and the gas-liquid interface, whose strength is set by the interfacial chemistry of the suspension. For a fixed particle material, varying $C_{in}$ therefore corresponds to varying the suspension chemistry at fixed particle size (e.g. pH).

\begin{figure}
  \centering
  \includegraphics[width=\linewidth]{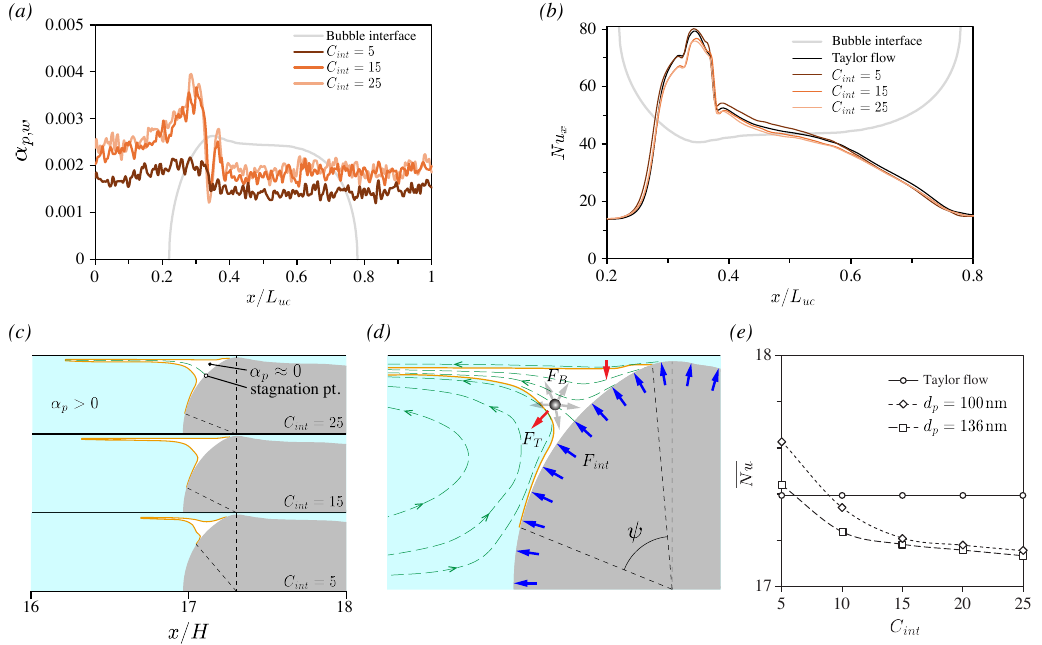}
  \caption{Effect of the particle-interface force coefficient on (\textit{a}) the near-wall nanoparticle volume fraction, (\textit{b}) the local Nusselt number distributions over a representative Taylor flow unit cell, (\textit{c,d}) the local particle volume fraction field at the rear of the bubble, and (\textit{e}) the wall-averaged Nusselt number for $d_p=100$ and $136~\mathrm{nm}$, with particle-free Taylor flow included as a reference. Results in (\textit{a-d}) are for a particle diameter, $d_p = 100~\mathrm{nm}$.}
  \label{fig:Cint_Nu_combined}
\end{figure}

Comparing the wall-averaged Nusselt number in figure~\ref{fig:Cint_Nu_combined}, a monotonic decrease is observed as the interfacial force ($C_{int}$) increases for both particle diameters. When $C_{int}=5$, both particle-laden cases exceed the particle-free Taylor flow reference, whereas stronger interface forcing reduces the thermal performance below that reference. The two diameters exhibit the same overall trend, although the $100~\mathrm{nm}$ particles produce the larger response. The remainder of this section therefore uses $d_p=100~\mathrm{nm}$ to isolate the underlying mechanism.

\begin{figure}
  \centering
  \includegraphics[width=\linewidth]{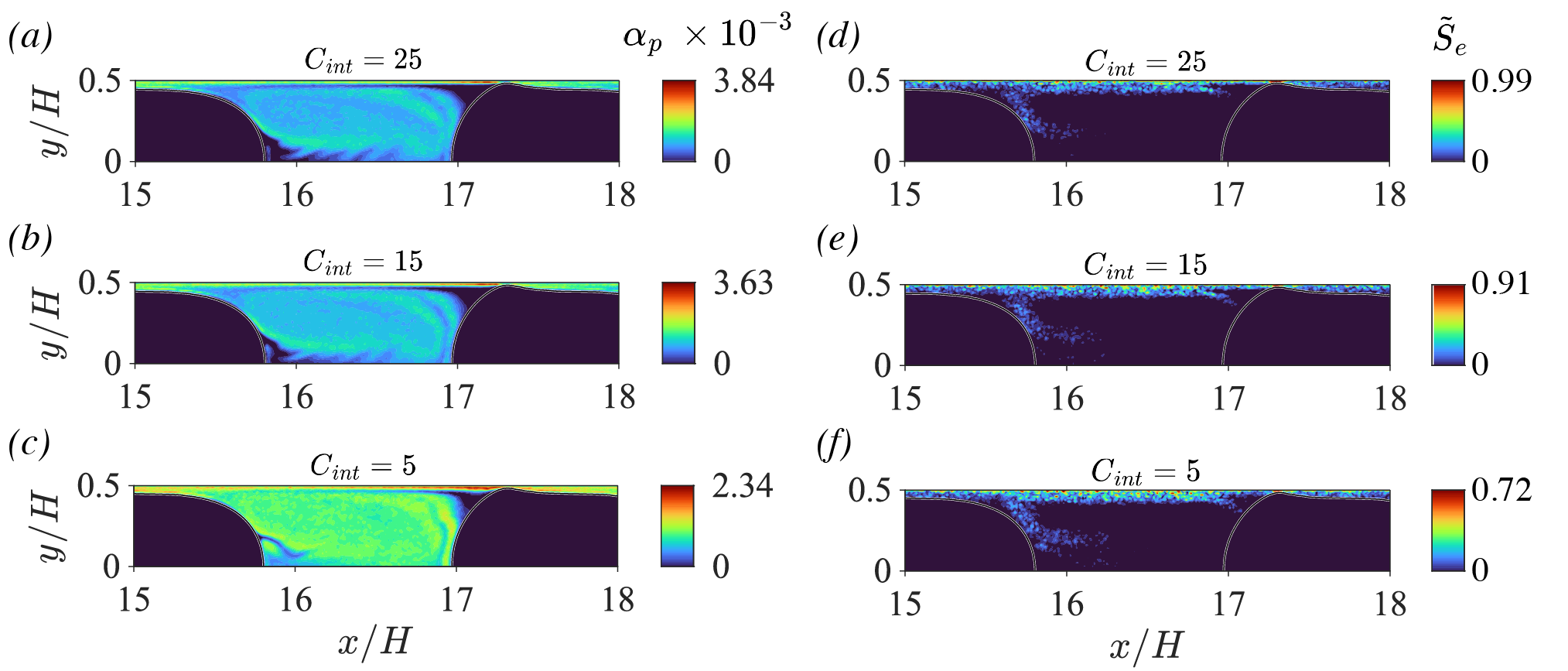}
  \caption{Effect of the particle-interface force coefficient on (\textit{a-c}) the spatial distribution of the particle volume fraction and (\textit{d-f}) the distribution of particle-fluid energy exchanges for $C_{int}=25$, $15$, and $5$.}
  \label{fig:Cint_fields}
\end{figure}

Examination of the particle volume fraction fields reveals a redistribution of the nanoparticles that is sensitive to the interface force (figure~\ref{fig:Cint_fields}\textit{a-c}). When $C_{int}\ge15$, particles are increasingly displaced from the gas-liquid interface, particularly around the bubble caps and within the thin liquid film. As the film layer is bounded by the interface on one side and the wall on the other, this stronger interface-side repulsion shifts a greater fraction of the particles toward the near-wall part of the film. By contrast, $C_{int}=5$ produces a more spatially balanced distribution, with particles remaining closer to the interface and more uniformly dispersed through the liquid slug.

This particle transport behaviour in the thin film is supported by the near-wall concentration profiles shown for a unit-cell (figure~\ref{fig:Cint_Nu_combined}\textit{a}). Here, the $C_{int}=15$ and $25$ cases give similar and generally higher near-wall concentrations, whereas $C_{int}=5$ gives a lower but more uniform profile along the unit-cell. The superior heat transfer response at lower interface forces, however, is not only due to the local particle concentration near the wall. Instead, we find that it is associated with the broader redistribution of particles between the interface, film, and recirculating slug (figure~\ref{fig:Cint_fields}\textit{a-c}). 

The primary reason that this short-range surface force can influence the particle distribution to such an extent within the film and liquid slug regions is due to the delicate balance between the hydrodynamic, thermal (Brownian), and particle-interface repulsion forces around the bubble end cap (figure \ref{fig:Cint_Nu_combined}\textit{c,d}). This region contains stagnant hydrodynamic conditions that separate the near-wall and internal slug streams, combined with high temperatures and temperature gradients (figure \ref{fig:istherms (section 5.1)}\textit{d}). As we have shown previously when investigating Brownian effects (Sec. \ref{sec:Brownian&thermophoretic}), this region where $\alpha_p\approx0$, emerges hydrodynamically from the stagnation point separating the convective circulations of the slug with the flow exiting the thin film around the bubble end cap. Particles exiting the thin film are influenced by the interface when this force is stronger than the Brownian and thermophoretic drift. The particles are deflected closer to the wall region and subsequently carried downstream in a high concentration near-wall flow ($C_{int} = 25$, figure \ref{fig:Cint_Nu_combined}\textit{c}, figure \ref{fig:Cint_Nu_combined}\textit{d}). Particles located below the stagnation point are forced into the slug flow circulation under the net influence of the interface force and thermophoresis (figure \ref{fig:Cint_Nu_combined}\textit{d}). As the flow moves downstream, the outcome is a stretched, or extended zone where $\alpha_p \approx0$, and there are poor particle momentum and energy exchanges between the near-wall and slug regions.

The angle the interface force influences particles along the bubble end cap is described by $\psi$ (figure \ref{fig:Cint_Nu_combined}\textit{d}). As the interface force reduces, this particle-separation angle also reduces from $\psi = 63.2^\circ$ at $C_{int} = 25$ to $\psi = 33.4^\circ$ at $C_{int} = 5$. Decreasing $C_{int}$ alters the balance of forces on the particle, with Brownian effects becoming dominant, reducing the size of the $\alpha_p\approx0$ region, and forcing particles from the near-wall and slug regions to move closer towards the stagnation point. This improves particle transport and energy exchanges downstream, shown in the corresponding distributions of $\tilde{S}_e$ (figure \ref{fig:Cint_fields}\textit{d-f}). These show that the high intensity near-wall thermal exchanges ($16 < x/H < 17$) extend further towards the channel centreline as the interface force reduces, changing from $|y/H| = 0.423$ ($C_{int} = 25$) to $|y/H|=0.399$ ($C_{int} = 5$). Through these mechanisms that redistribute the particles, and improve thermal exchanges between the near-wall and slug flow regions, the local Nusselt number is enhanced above that of particle-free Taylor flow (figure \ref{fig:Cint_Nu_combined}\textit{b}). Collectively, these results show how short-range particle interactions at the interface can introduce notable effects on wider transport phenomena within Taylor flows containing nanofluids.

\subsection{Flow structure effects: Reynolds number, slug length and void fraction}
\label{sec:flow_structure_effects}

The effects of Reynolds number, unit-cell length, and void fraction were examined systematically using nanofluid suspensions with $d_p=78~\mathrm{nm}$, the smallest particle diameter that produced the largest $\eta_{Nu}$ at $\phi=10^{-3}$. The Reynolds number was varied over $500\leq Re\leq1000$ at $\varepsilon=0.5$ and $L_{uc}/H=2.66$. The unit-cell length was subsequently varied over $2.66 \leq L_{uc}/H \leq 5$ at $Re=1000$ and $\varepsilon=0.5$, while the void fraction study considered $\varepsilon=0.33$, $0.50$, and $0.66$ at $Re=1000$ and $L_{uc}/H=4$, respectively.

\begin{figure}
  \centering
  \includegraphics[width=1.0\textwidth]{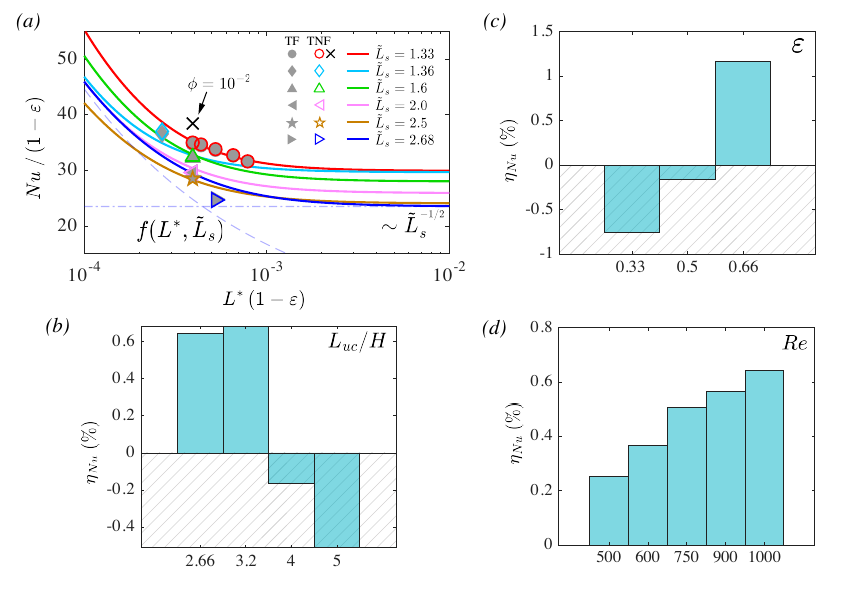}
  \caption{Effect of Taylor flow parameters on the wall-averaged Nusselt number. (\textit{a}) Nusselt number predictions for all parameters plotted against thermal entrance length, $L^* = L_{th}/D_hRePr$, and accounting for the presence of the gas phase. Data are presented as dimensionless slug length, $\tilde{L}_s = L_s/H$, and compared to a semi-empirical model adapted from \cite{Walsh2010} for constant heat flux circular tubes. Note that $Nu$ from the semi-empirical model are not modified by $(1-\varepsilon)$ on the ordinate of (\textit{a}), only the numerical simulation data. Enhancement to wall-averaged Nusselt number over particle-free Taylor flow for various (\textit{b}) normalised unit-cell lengths, $L_{uc}/H$, (\textit{c}) void fraction, $\varepsilon$, and (\textit{d}) Reynolds number, $Re$. }
  \label{fig:Nu_all}
\end{figure}

The predictions across these parameter spaces have been consolidated into a single dimensionless heat transport plot by representing the data in terms of the dimensionless liquid slug length, $\tilde{L}_s = L_s/H$, and accounting for the effects of the gas phase fraction, $\varepsilon$, on the Nusselt number and thermal entrance length, $L^* = L_{th}/D_hRePr$ (figure \ref{fig:Nu_all}\textit{a}). The simulations have been plotted against a semi-empirical model by \cite{Walsh2010} that has been adapted for the constant heat flux channel problem investigated in this work. Here, two asymptotic limits are considered which combine the effects of the developing and fully developed flow regions in the form, $Nu = (Nu_d^{^{7/2}}+Nu_{fd}^{^{7/2}})^{^{2/7}}$, using the blending approach of \cite{ChurchillOzoe1973}. The analytical solutions for Poiseuille flow ($Nu\sim L^{*^{-1/3}}$) and for plug flow ($Nu\sim L^{*^{-1/2}}$) in the channel entrance are combined using a weighted mean that biases the solution to plug-flow-like behaviour in short slugs (e.g. $\tilde{L}_s\approx 1$), and Poiseuille-flow-like behaviour in the long-slug limit ($\tilde{L}_s\gg 1$) \citep{Muzychka2010,Walsh2010}. This leads to a description of the Nusselt number in the developing flow region, $Nu_d = 1.49/L^{*^{1/3}} + \tilde{L}_s^{^{-1}}(0.886/L^{*^{1/2}}-1.49/L^{*^{1/3}})$. The fully developed region has been suggested to scale according to a combination of the Poiseuille flow solution and the contributions of flow segmentation effects, empirically determined to be $25\tilde{L}_s^{^{-1/2}}$ for a constant heat flux circular channel \citep{Walsh2010}. Extending this to our flow configuration, where $Nu = 8.235$ for the Poiseuille flow solution (figure \ref{fig:validation_all}\textit{b}), we use $Nu_{fd} = 8.235+25\tilde{L}_s^{^{-1/2}}$. An example of the two asymptotes describing the developing flow limit, where $Nu = f(L^*,\tilde{L}_s)$, and the fully developed flow limit, where $Nu\sim \tilde{L}_s^{^{-1/2}}$, are shown in figure \ref{fig:Nu_all}\textit{a} for $\tilde{L}_s = 2.68$. 

Overall, the results demonstrate very good agreement between the semi-empirical model and the numerical simulations, indicating that the scaling between slug length, plug flow, and Poiseuille flow is a reasonable representation of the thermal and hydrodynamic behaviour. The predictions for Taylor nanofluid flow (TNF) at low nanoparticle concentrations, $\phi = 10^{-3}$, are almost indistinguishable from particle-free Taylor flows (TF) when viewed across these wide hydrodynamic conditions, $L^*(1-\varepsilon)\sim10^{-4}-10^{-2}$. This confirms that the liquid-gas flow topology remains mostly unaltered by the presence of the nanoparticles (figure \ref{fig:Nu_v_dp_phi}\textit{b}). However, when the particle volume fraction is increased to $\phi = 10^{-2}$, we see a notable departure from the expected heat transport for particle-free Taylor flow and an equivalent slug length. This highlights a threshold where the addition of nanoparticles have been found to amplify non-dimensional heat transport by $\approx10\%$ for $d_p=78~\mathrm{nm}$, accompanied by significant alterations in Taylor bubble topology and film thickness (figure \ref{fig:Nu_v_dp_phi}\textit{a,b}).

The predictions for varying unit-cell length ($L_{uc}/H$), void fraction ($\varepsilon$), and Reynolds number ($Re$) have been isolated and replotted in terms of $\eta_{Nu}$ to identify trends underlying the subtle changes in $Nu$ observed at this low particle volume fraction (figure \ref{fig:Nu_all}\textit{b-d}). For varying unit-cell length and fixed $\varepsilon = 0.5$, this reveals a more complex relationship, where only the smaller unit-cells show positive enhancement over particle-free Taylor flows (figure \ref{fig:Nu_all}\textit{b}). This trend is not exclusive to $L_{uc}/H$, however, and depends on the void fraction. For variable void fraction and a fixed unit-cell length of $L_{uc}/H=4$, enhancements to $Nu$ can become positive when $\varepsilon > 0.5$ (figure \ref{fig:Nu_all}\textit{c}). Collectively, this indicates short slug lengths of $\tilde{L}_s \leq 1.6$ and nanoparticle addition creates mutually beneficial heat transport phenomena that leads to an increase in the wall-averaged Nusselt number.

\begin{figure}
  \centering
  \includegraphics[width=\linewidth]{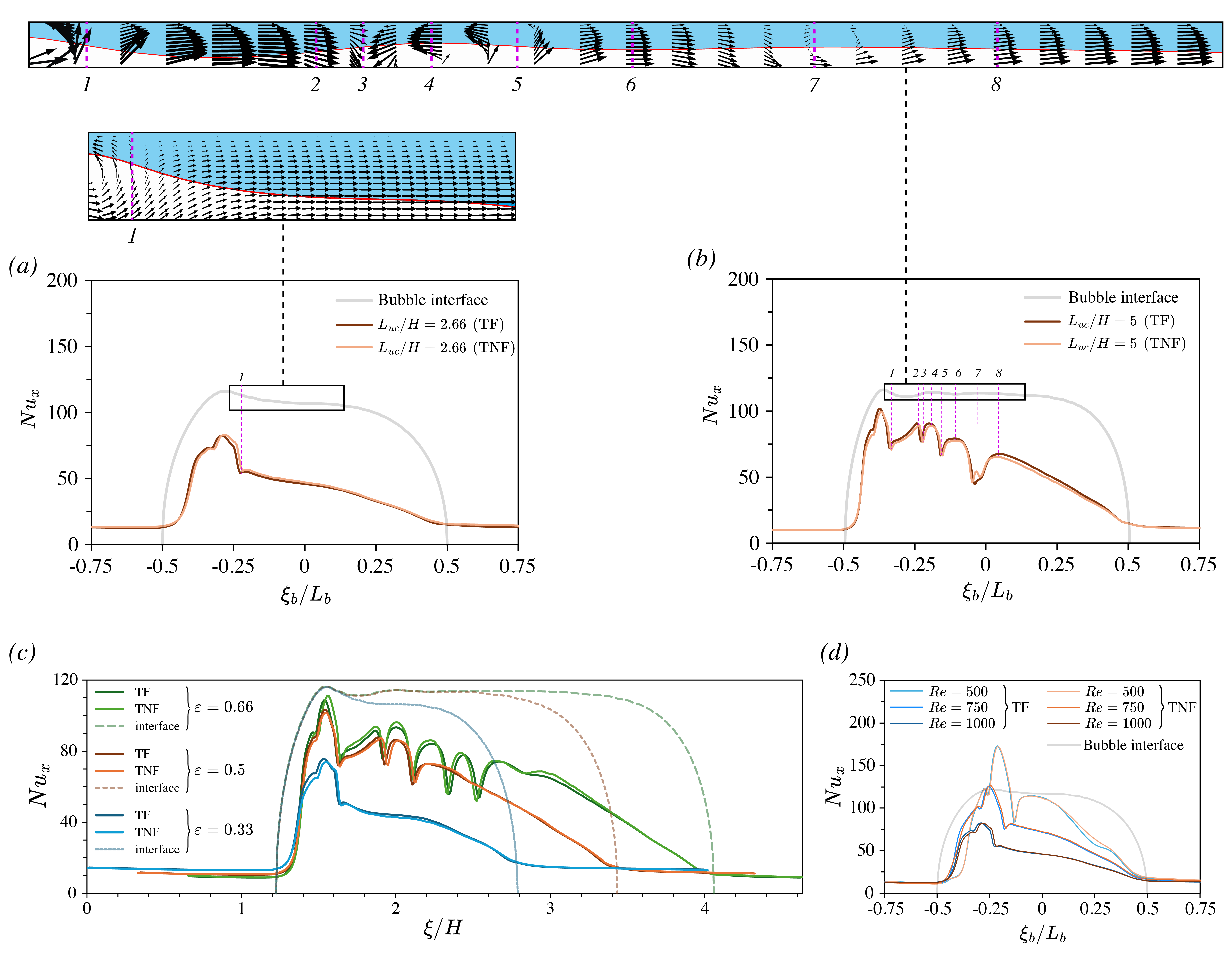}
  \caption{Effect of Taylor flow parameters on the local Nusselt number distribution, showing variations in (\textit{a,b}) unit-cell length and the corresponding velocity structure within the liquid film region for $L_{uc}/H=2.66,~5$, (\textit{c}) void fraction for $\varepsilon=0.33$, $0.50$, and $0.66$, and (\textit{d}) Reynolds number, $Re = 500$, $750$, and $1000$. Results are shown for particle-free Taylor flow (TF) and particle-laden Taylor flow (TNF), with the bubble interface curve identifying the bubble region. Snapshots of the flow fields in the film region are presented for the velocity, $\tilde{\mathbf{u}}$, in the laboratory frame of reference.}
  \label{fig:Nu_Luc_local}
\end{figure}

To understand the origins of this phenomena, we examined the local Nusselt number distributions along the thin film and slug flow regions (figure \ref{fig:Nu_Luc_local}). For a constant void fraction, $\varepsilon = 0.5$, the short unit-cell length predominantly increases $Nu_x$ along both the film and slug regions of the unit-cell (figure \ref{fig:Nu_Luc_local}\textit{a,b}). This feature is reversed when $L_{uc}/H = 5$, as the Taylor nanofluid flow reduces $Nu_x$ along the unit-cell. The local film flow structure provides further insight into the $Nu_x$ distributions. The shorter unit cell exhibits a single dominant velocity reversal within the film with a comparatively smooth local Nusselt number distribution. By contrast, the longer unit cell contains several alternating flow structures, which coincide with the successive oscillations in $Nu_x$. Here, the the local $Nu_x$ minima along the film coincide with the boundaries of recirculation zones, where the axial velocity approaches zero, a similar behaviour as reported by \citet{Gupta2010}. Thus, successive film vortices produce corresponding oscillations in the local heat transfer response. 

However, the presence of these successive film vortices is not the reason for local reductions in $Nu_x$ for particle-laden Taylor flows. By varying the void fraction and maintaining a constant unit-cell length, $L_{uc}/H = 4$, we observe the opposite trend to the previous investigation where void fraction was fixed, and unit-cell length varied (figure \ref{fig:Nu_Luc_local}\textit{c}). In this scenario, the longer film with multiple $Nu_x$ oscillations locally increases when nanoparticles are added ($\varepsilon = 0.66$), whereas heat transport for the shorter film ($\varepsilon = 0.33$) locally reduces. The situation is clarified when considering all Taylor flow parameters investigated in figure \ref{fig:Nu_Luc_local}. We find that the liquid slug length, $\tilde{L}_s$, is the key parameter that governs whether positive changes occur for \textit{both} the local film and slug heat transfer. Interestingly, the mutual benefits of maintaining $L_s^{+} \leq 1.6$ when adding nanoparticles holds for both the film and slug regions, despite the hydrodynamic development of the thin film surrounding Taylor bubbles having no sensitivity to this parameter \citep{HAN2009}. Instead, it is the intensity of the convective mixing within the slug which influences particle momentum and energy exchanges from the near-wall region, indirectly improving the particle distribution in the film for successive Taylor bubbles. As slug length is reduced, transport phenomena approach that of a plug-flow-like behaviour, redistributing particles efficiently throughout the liquid carrier phase, whereas longer slugs have lower convective mixing intensity that weakens sufficiently when $L_s^{+} > 2$, leading to a reduction in $Nu$ far before the Poiseuille flow limit, $\tilde{L}_s\gg100$.         

Further examination of the effects of varying $Re$ on $\overline{Nu}$ and $Nu_x$ confirm this (figure \ref{fig:Nu_all}\textit{d}, figure \ref{fig:Nu_Luc_local}\textit{d}). Increasing $Re$ also plays a role in increasing the convective transport within the slug, and pushing the channel flow to smaller thermal entrance lengths, $L^*$. Although the absolute local $Nu_x$ in the film is highest for the low Reynolds number due to a reduction in inertial effects, $Ca$, and subsequently $\delta$, the relative change to $Nu$ between Taylor flow and particle-laden Taylor flow monotonically increases with increasing $Re$. This enhancement occurs for the same reason as reducing slug length -- improved convective mixing and nanoparticle transport from the near-wall region to the bulk nanofluid. These results show that nanoparticle addition in Taylor flows does not provide a uniform enhancement to non-dimensional heat transport. Instead, its benefit increases when the underlying flow structure promotes sufficiently strong recirculation and repeated redistribution of the particles and their associated thermal energy exchange. 

\section{Conclusions}
\label{sec:conclusions}
In this study, we investigate nanoparticle transport in non-isothermal, gas-liquid Taylor flow. A fully coupled Eulerian--Lagrangian framework has been developed that combines a volume of fluid description of the gas-liquid phases with a multi-phase particle-in-cell representation of the dispersed nanoparticles, accounting explicitly for particle-fluid momentum and energy exchange, particle interaction with the gas-liquid interface, Brownian motion, and the resulting thermophoretic drift. This model was verified against Brownian-diffusion and liquid-phase thermophoresis benchmarks, and validated against independent investigations on Taylor flow and $\mathrm{Al_2O_3}$--water nanofluid suspensions.

Using this model, we systematically investigated the role of hydrodynamic, thermal, and interfacial forces on nanoparticle transport, particle-fluid energy exchanges, and non-dimensional heat transport phenomena for particle diameters, $d_p\sim10-100~\mathrm{nm}$, and particle volume fraction, $\phi\sim10^{-3}-10^{-2}$. We find that nanoparticles can enhance the wall-averaged Nusselt number in Taylor bubble flows by almost $10\%$ in this parameter range and for the visco-inertial regime ($Re>500$). However, a net reduction over classical Taylor flows was also observed and only avoided for the largest particles when $\phi = 10^{-2}$. We find that the interplay between the hydrodynamics and Brownian effects play a key role on whether the addition of nanoparticles increases, or decreases, local and wall-averaged dimensionless heat transport. At the highest concentration and smallest particle diameter, Brownian motion promotes particle redistribution and energy exchanges, while the bubble topology is altered considerably, reducing the film thickness around the Taylor bubble by over 50\%, and significantly enhancing local Nusselt number. 

The addition of nanoparticles to the liquid phase was also found to alter transport phenomena differently for Poiseuille flows and Taylor bubble flows in confined, constant heat flux channels. Poiseuille nanofluid flows are governed by the competing effects of stochastic Brownian motion and thermophoretic drift. The latter transports nanoparticles away from the boundary layer in the heated near-wall region, reducing the Nusselt number compared to non-Brownian, particle-laden Poiseuille flows. Taylor bubble flows containing nanofluids present contrasting behaviour. Here, Brownian motion supports the redistribution of particles from the film and near-wall regions throughout the liquid phase, and the convective motion from the slug circulations opposes the adverse effect of thermophoretic particle depletion from the near-wall. Interestingly, short-range particle-interface interactions, most notably around the bubble end-cap, were found to influence the particle distribution within the wider film and slug regions. The origin of this effect was located near the film exit at the rear of the bubble, and around the stagnation point at the interface. As the interface force increased, a low particle concentration wake region behind the bubble extended further downstream, creating a barrier to nanoparticle mixing between the slug circulations and the near-wall flow and influencing the concentration in the thin film around the bubble. Lower particle repulsion energies from the bubble interface subsequently enhanced heat transport phenomena. As many of the established Eulerian formulations and experimental data reduction strategies account for the presence of nanoparticles in liquids using models describing effective thermophysical properties, this work demonstrates that these approaches may have limited applicability due to the strongly coupled multi-phase transport phenomena observed across this wide disparity of scales.

There is a sensitive connection between particle transport within the convective liquid slug circulations, and the resulting particle transport and heat and mass transfer behaviour of the thin film surrounding the bubble. Progressive increases in local and wall-averaged Nusselt number over particle-free Taylor flow emerge when the dimensionless slug length, $\tilde{L}_s \leq 1.6$. This is not evident in classical Taylor bubble flows in the visco-inertial regime explored here, where thin film characteristics are only a function of the Reynolds and Capillary numbers, and independent of $\tilde{L}_s$. Shorter slugs or increasing Reynolds number, were found to strengthen nanoparticle mixing and transport between the near-wall region and channel centreline. This benefits particle transport and energy exchanges along both the liquid slug and the thin film regions around successive Taylor bubbles. It also provides the first mechanistic explanation for the progressively increasing heat transport with decreasing slug length recently observed experimentally \citep{Alrbee2022}.  

The notable changes in ${Nu}$ and bubble topology as nanoparticle concentrations reach $\phi = 10^{-2}$ indicate that future explorations of denser particle regimes up to $\phi = 10^{-1}$ could provide valuable insights. The simplifications assumed here for particle-interface interactions supported this investigation, but is a limitation that also warrants further model development, including capturing particle attraction and repulsion forces that change with interface proximity and surface chemistry \citep{Bau2026}. Similarly, although temperature-dependent thermophysical properties have been implemented in this work, and surface tension was found to vary along the interface by $<1\%$, future investigations could consider examining the effects of Marangoni stresses due to surface tension gradients. This may be particularly relevant for nanofluid dispersions that are stabilised in the liquid phase using surfactants. Furthermore, given the motivation of our work is orientated toward obtaining an improved understanding of transport phenomena in microfluidic and advanced cooling applications that involve nanofluids, our numerical method has the potential to be extended for investigating problems including temperature-dependent turbulence modulation in forced convection \citep{Zonta_Marchioli_Soldati_2012}, low Reynolds number flow boiling and condensation \citep{McGrail2013}, combined particle and solute transport around Taylor bubbles \citep{Picchi_Poesio_2022}, and Taylor bubble pinch-off in complex fluids \citep{Richards_Pegler_2026}.

\section*{Acknowledgements}
The computations described in this paper were performed using the University of Birmingham's BlueBEAR HPC service, which provides a High Performance Computing service to the University's research community.

\section*{Funding}
This research was supported by the EPSRC research grant [EP/Z002540/1], the Higher Committee for Education Development in Iraq (HCED) Scholarship program, and a Royal Academy of Engineering/Leverhulme Trust Research Fellowship [LTRF2122-18-108].

\section*{Declaration of interests}
The authors report no conflict of interest.

\appendix

\section{Solution methods and numerical schemes}\label{appA}
\begin{figure}
\centerline{\includegraphics[scale=0.7]{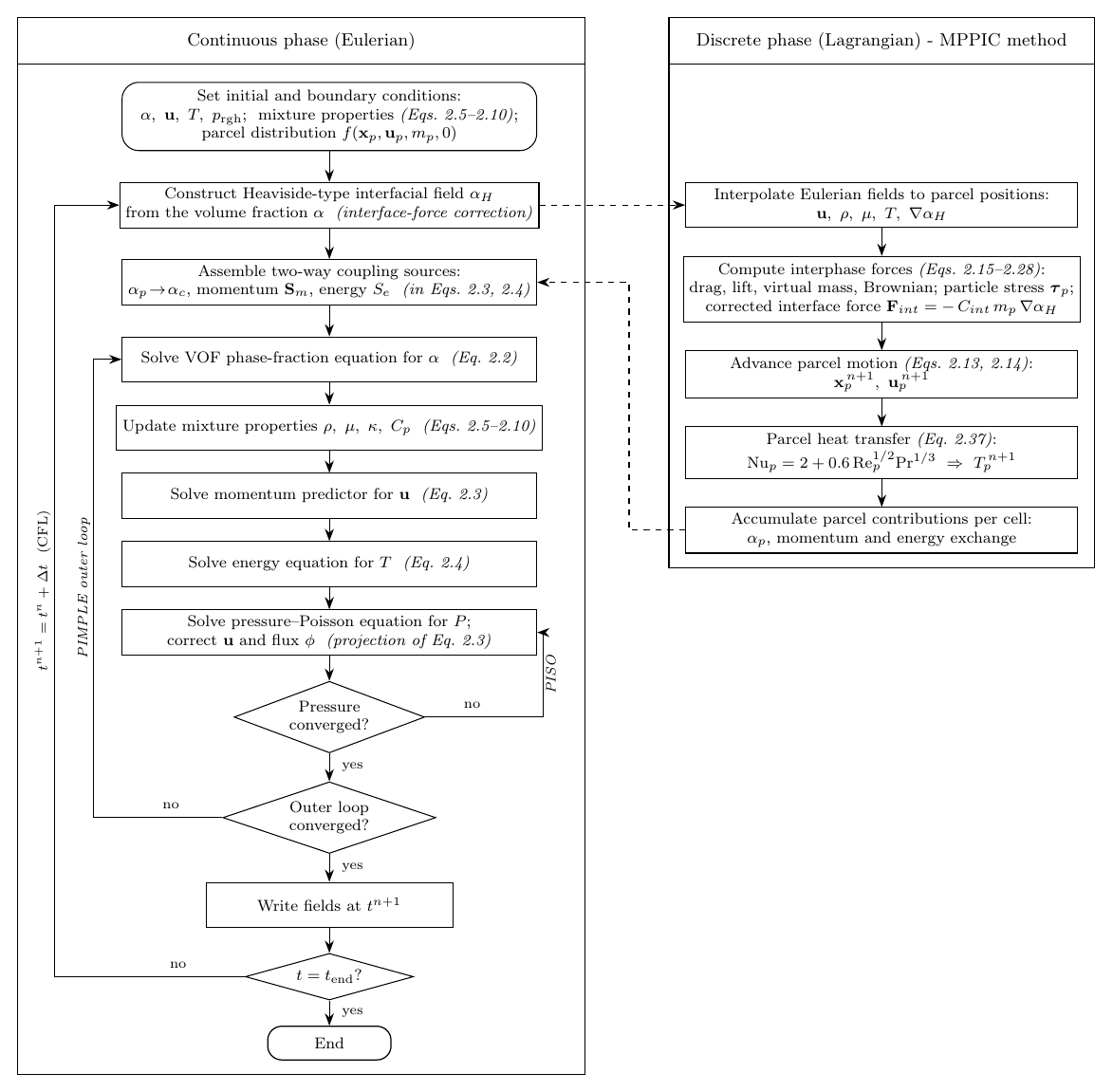}}
  \caption{Flow chart illustrating the Eulerian-Lagrangian solver developed for this study.}
\label{fig:flow chart}
\end{figure}

\noindent The coupled gas-liquid-particle momentum and energy equations were solved using a customised solver developed within the OpenFOAM-v2312 finite volume framework. The in-built \texttt{MPPICInterFOAM} solver was used as a starting point and substantially modified and extended to form a new solver, \texttt{thermalMPPICInterFOAM} (figure \ref{fig:flow chart}). This enabled the solution of the energy equation in the Eulerian phase together with Eulerian-Lagrangian thermal coupling for the dispersed particle phase. Additional phase-property submodels were incorporated to account for temperature-dependent and phase-dependent thermophysical properties, including density, viscosity, heat capacity, and thermal conductivity. The pressure--velocity coupling was handled using a PIMPLE-based algorithm with one outer corrector and three pressure correctors. Spatial interpolation of Eulerian fields to Lagrangian particle locations was performed using both \texttt{cell} and \texttt{cellPoint} interpolation schemes. The \texttt{cell} scheme was used for scalar thermophysical properties such as $\rho$, $\mu$, $c_p$, $\kappa$, and $T$, while the \texttt{cellPoint} scheme was used for fields with stronger spatial gradients, including $\mathbf{u}$, $\nabla T$, $\nabla \alpha$, $D\mathbf{u}_c/Dt$, and $\nabla \times \mathbf{u}_c$. Particle momentum and energy equations were integrated analytically, and the corresponding source terms for $\mathbf{u}$ and $T$ were treated semi-implicitly. The solution methods and convergence criteria are summarised in Table~\ref{tab:solution}, while the discretisation schemes are listed in Table~\ref{tab:scheme}. To maintain numerical stability, the maximum Courant number was limited to 0.2 for the coupled fluid-particle calculation.

\begin{figure}
\centering
\includegraphics[width=0.6\textwidth]{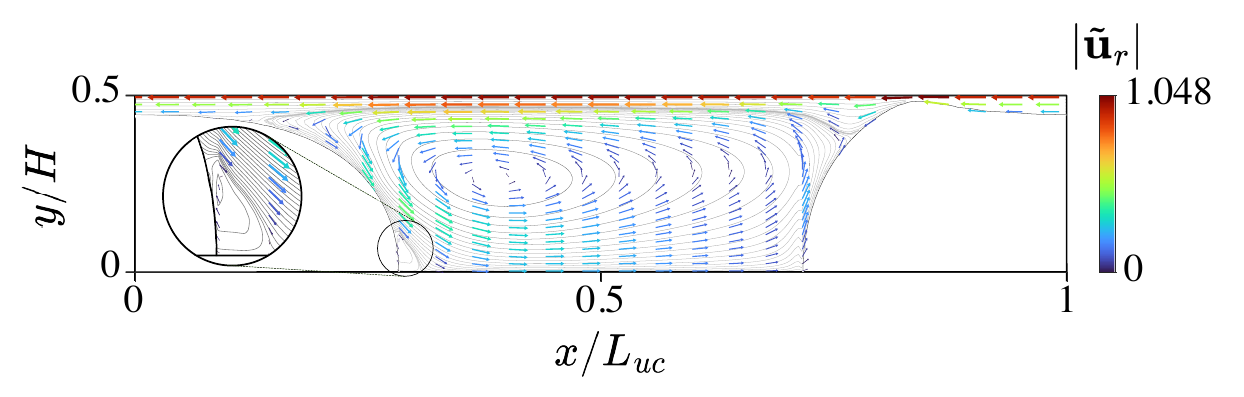}
\caption{Instantaneous velocity field within a Taylor-flow unit cell, showing the scale of a residual spurious current near the gas-liquid interface.}
\label{fig:currents}
\end{figure}

The gas-liquid interface was captured using the MULES-based Volume-of-Fluid formulation with two alpha sub-cycles and an interface-compression coefficient of $c_\alpha=0.7$, following \citet{Larsen2019}. In this context, the last term in equation~(\ref{Alpha transport}) represents the artificial compression term used to maintain a sharp gas-liquid interface. This treatment was combined with a symmetric Gauss--Seidel smoother using 100 iterations, which improved interface sharpness, enhanced numerical stability, helped manage the large property gradients across the compressed interface, and reduced spurious currents \citep{ABADIE2015}. As shown in figure \ref{fig:currents}, spurious currents around the bubble interface were minimal, and mainly appeared ahead of the bubbles, away from channel wall, where their influence on the wall heat transport was limited.

A High-Performance Computer (BlueBEAR), configured with six water-cooled nodes featuring Ice Lake CPUs, was utilized for parallel computations using the OpenMPI implementation of the Message Passing Interface (MPI). Each node was equipped with two Intel\textregistered{} Xeon\textregistered{} Platinum 8360Y processors, each with 36 cores and 512 GB of RAM. The simulation wall time varied according to the number of parcels and the prescribed end time, whereas the number of Eulerian cells was kept constant. In the present study, the largest simulation involved 32 million parcels. In most cases, each parcel represented 10 particles; however, for a limited number of cases with smaller particle diameters, this value was increased to 50 particles per parcel. An analysis was performed on the effects of particle-per-parcel which informed these settings (figure \ref{fig:hindered_PPP}). Details of the mesh sensitivity and cell-size-to-particle-size sensitivity analyses are provided in Appendix \ref{appD} below. 

\begin{table}
  \begin{center}
  \def~{\hphantom{0}}
  \begin{tabular}{lll}
      Variable & Solution method & Convergence \\[3pt]
      $\alpha$ & smoothSolver + symGaussSeidel & $10^{-9}$ \\
      $\alpha_c$ & GAMG + GaussSeidel & $10^{-7}$ \\
      $P$ & PCG + DIC & $10^{-8}$ \\
      $\mathbf{u}$ & smoothSolver + symGaussSeidel & $10^{-7}$ \\
      $T$ & PBiCGStab + DILU & $10^{-7}$ \\
  \end{tabular}
  \caption{Linear controls and convergence criteria used in \texttt{thermalMPPICInterFOAM}. PCG: Preconditioned Conjugate Gradient; DIC: Diagonal Incomplete Cholesky Preconditioner; DILU: Diagonal Incomplete LU; GAMG: Geometric Algebraic Multi-Grid; PBiCGStab: Preconditioned Biconjugate Gradient Stabilised method}
  \label{tab:solution}
  \end{center}
\end{table}

\begin{table}
  \begin{center}
  \def~{\hphantom{0}}
  \begin{tabular}{llll}
      Category & Symbol & Numerical scheme & Remarks \\[3pt]
      Temporal & $\partial/\partial t$ & Euler & first-order implicit \\
      Gradient & $\nabla$ & Gauss linear & second-order \\
      Divergence (convection) & $\nabla \cdot (\alpha_c \rho \mathbf{u} \mathbf{u})$ & Gauss upwind & first-order upwind scheme \\
      & $\nabla \cdot \left(\alpha_c \mathbf{u}\right)$ & Gauss vanLeer & second-order accuracy \\
      & $\nabla \cdot \left(\rho c_p {T}\mathbf{u}\right)$ & Gauss vanLeer & second-order accuracy \\
      & $\nabla \cdot \left(\alpha_c \alpha \mathbf{u}\right)$ & Gauss linear & second-order accuracy \\
      & $\nabla \cdot \left(\alpha_c \alpha (1- \alpha) \mathbf{u_c}\right)$ & MULES & second order semi-implicit \\
      Diffusion & $\nabla^2$ & Gauss linear corrected & non-orthogonal correction \\
  \end{tabular}
  \caption{Numerical discretisation schemes used for the system of equations.}
  \label{tab:scheme}
  \end{center}
\end{table}

\section{A simple model of the net interface force}\label{appB}

\begin{figure}
\centering
\includegraphics[width=\textwidth]{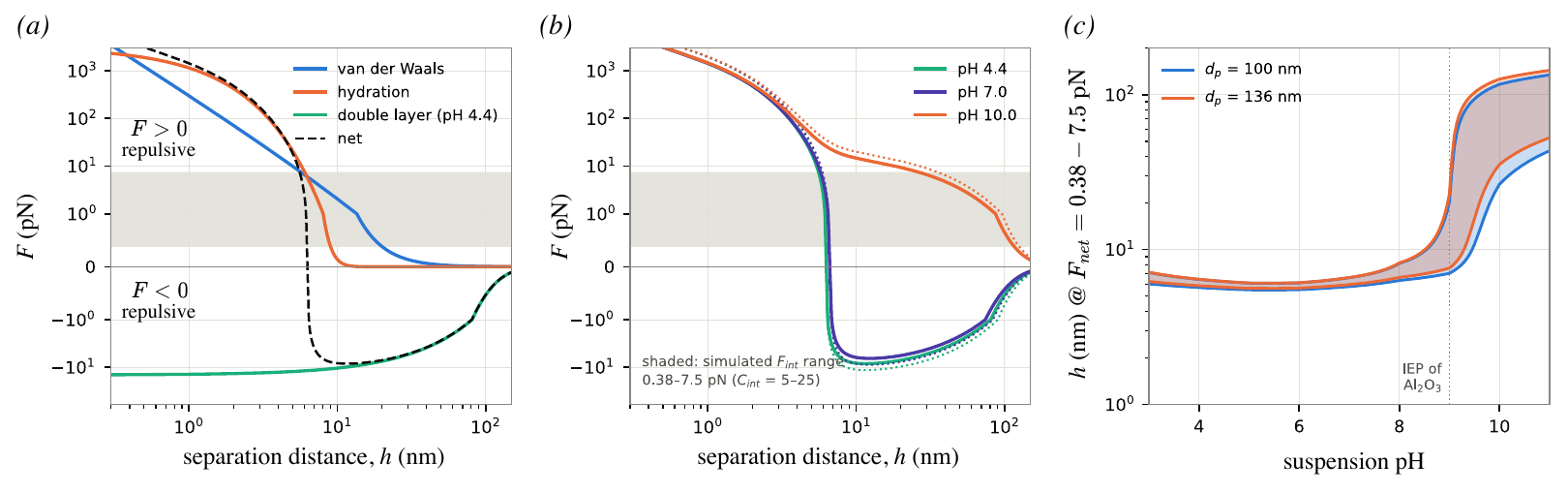}
\caption{Net surface force on an Al$_2$O$_3$ nanoparticle approaching an air--water interface ($I = 10^{-4}$~M). (\textit{a})~Components and net force for $d_p = 100$~nm at pH~4.4. (\textit{b})~Net force at pH~4.4, 7 and 10 for $d_p = 100$~nm (solid) and 136~nm (dotted); the shaded band is the range of interface force per particle applied in the simulations, 0.38--7.5~pN, for $C_{int} = 5$--25. (\textit{c})~Separation at which the net repulsion equals the simulated range, as a function of pH; the band spans 0.38~pN (upper curve) to 7.5~pN (lower curve).}
\label{fig:netforce}
\end{figure}

The interface force applied across the numerical simulations was compared to a simple model describing the net force due to the approach of an Al$_2$O$_3$ nanoparticle to a water-air interface (eq. \ref{eq:net_int_F}). A realistic range of interface force coefficients, $C_{int}$, were determined by extracting the force per particle along the bubble interface and comparing to these model predictions. This is described here for the four bounding cases examined in Sec. \ref{sec:interface_coefficient} ($d_p = 100$ and 136~nm, $ C_{int}= 5$ and 25). Since $|\mathbf{F}_{int}| = m_p C_{int} |\nabla\alpha_H|$ and $|\nabla\alpha_H|$ lies between $3.5\times10^{4}$ and $5.8\times10^{4}~\mathrm{m^{-1}}$ along the interface (i.e. the numerical interface has an effective thickness $1/|\nabla\alpha_H| \approx 17$--$28~\mathrm{\mu m}$), the force per particle spanned 0.38--7.5~pN (Table~\ref{tab:hrange}). The separation distance, $h$, from a real water-air interface where an Al$_2$O$_3$ particle experiences a net repulsion of this magnitude is presented in figure \ref{fig:netforce}. This also shows how that separation distance depends on the pH of the suspension. The comparison is between force magnitudes, and does not compare force profiles, since the physical forces act over nanometres and the model force in the simulations is over the numerical interface thickness.

\subsection{Model}

The net force is the sum of the three contributions described in Sec. \ref{sec:formulation} (eq. \ref{eq:net_int_F}), and evaluated for a sphere of radius $R = d_p/2$ and a planar interface (the bubble radius, $\approx 0.5$~mm, is four orders of magnitude larger than $R$), with $F>0$ repulsive:

\begin{equation}
  F(h) = F_{\mathrm{vdW}}(h) + F_{\mathrm{EDL}}(h) + F_{\mathrm{hyd}}(h).
  \label{eq:Fnet}
\end{equation}

\noindent For the van der Waals term the retarded form of \citet{Gregory1981} is used, since the separations of interest exceed a few nanometres,

\begin{equation}
  W_{\mathrm{vdW}}(h) = -\frac{A_{132}R}{6h}\left[1 - \frac{b h}{\lambda}\ln\!\left(1 + \frac{\lambda}{b h}\right)\right],
  \qquad b = 5.32,\; \lambda = 100~\mathrm{nm},
  \label{eq:vdWret}
\end{equation}

\noindent with $A_{132} = -3.8\times10^{-20}~\mathrm{J}$ and $F_{\mathrm{vdW}} = -d W_{\mathrm{vdW}}/d h$. For the electric double-layer term the linear-superposition approximation with effective potentials is used \citep{Bell1970,NguyenSchulze2004}:

\begin{equation}
  F_{\mathrm{EDL}}(h) = 64\pi\varepsilon_r\varepsilon_0 R\,\kappa \left(\frac{k_BT}{e}\right)^{2}
  \tanh\!\left(\frac{e\psi_1}{4k_BT}\right)\tanh\!\left(\frac{e\psi_2}{4k_BT}\right) e^{-\kappa h},
  \label{eq:LSA}
\end{equation}

\noindent which is accurate for $\kappa h \gtrsim 1$ and is an extrapolation closer in. Hydrated, hydroxylated oxide surfaces bind water strongly; removing that water to make solid-gas contact costs energy, which appears as a repulsion that decays exponentially with a characteristic length $\lambda_h$ \citep{IsraelachviliPashley1983,Israelachvili2011}. In the wetting-film literature this is the structural component of the disjoining pressure \citep{DerjaguinChuraev1974}. With the plate-plate form $W_{\mathrm{A}} = K_h e^{-h/\lambda_h}$, the sphere-plane energy and force are:

\begin{equation}
  W_{\mathrm{hyd}}(h) = 2\pi R K_h \lambda_h\, e^{-h/\lambda_h}, \qquad
  F_{\mathrm{hyd}}(h) = 2\pi R K_h\, e^{-h/\lambda_h},
  \label{eq:hyd}
\end{equation}

\noindent with $K_h \approx 3$--$30~\mathrm{mJ\,m^{-2}}$ and $\lambda_h \approx 0.6$--$1.1~\mathrm{nm}$ for hydrophilic surfaces \citep[ch.~15]{Israelachvili2011}. Here, we use $K_h = 10~\mathrm{mJ\,m^{-2}}$ and $\lambda_h = 1~\mathrm{nm}$. The hydration term is present for any hydrophilic solid, is insensitive to pH, and is what finally stops a hydrophilic particle from touching the interface even when the double-layer term is attractive.

The surface potentials are set equal to the zeta potentials of Al$_2$O$_3$ and of the water--air interface at the given pH. For Al$_2$O$_3$ the values measured for Al$_2$O$_3$--water nanofluids by \citet{AlMuhaysh2025} are used ($+50$, $+52$, $+25$, $0$ and $-33$~mV at pH 4, 6, 7, 9 and 10), with the isoelectric point at pH~$\approx$~9 \citep{Kosmulski2021}. For the interface, values representative of the published range are used ($-18$, $-32$, $-38$, $-55$ and $-70$~mV at the same pH values); the interface is negatively charged above an isoelectric point reported between pH~2 and~4.5 \citep{Takahashi2005,Beattie2009}, so that its potential at the experimental pH of 4.4 lies between about 0 and $-20$~mV, and the sensitivity of the results to this value is noted below. The base-case ionic strength is $I = 10^{-4}~\mathrm{M}$, representative of acid-adjusted deionised water without added electrolyte ($\kappa^{-1} = 30$~nm); $T = 303$~K.

\subsection{Results}

The outcomes from the model above are shown in figure~\ref{fig:netforce} and table~\ref{tab:hrange}. Below the isoelectric point of Al$_2$O$_3$ (pH~$\lesssim 8$), a net repulsion of 0.38--7.5~pN is reached at $h \approx 5.8$--6.5~nm, essentially independent of pH, ionic strength and particle diameter (figure~\ref{fig:netforce}\textit{c}). In this regime, the repulsion is set by the hydration and van der Waals terms, both of which rise steeply below about 8~nm. The position of this repulsive wall depends on the hydration parameters, which are the least certain input to the model: over $K_h = 3$--$30~\mathrm{mJ\,m^{-2}}$ and $\lambda_h = 0.6$--$1.1$~nm it moves between about 4 and 8~nm. Above the isoelectric point (pH~10), all three terms are repulsive and the same force range is reached at $h \approx 49$--131~nm for $I = 10^{-4}$~M (24--53~nm for $I = 10^{-3}$~M). This is due to the double-layer repulsion, with its Debye-length range, dominating at these conditions.

\begin{table}
  \begin{center}
  \def~{\hphantom{0}}
  \begin{tabular}{llcccc}
    $d_p$ (nm) & $C_{int}$ & $F_{\mathrm{int}}$ per particle (pN) & pH 4.4 & pH 7 & pH 10 \\
    100 & 5  & 0.38--0.61 & 6.3 & 6.4--6.5 & 116--131 \\
    100 & 25 & 1.89--3.06 & 6.0--6.1 & 6.1--6.3 & 67--82 \\
    136 & 5  & 0.93--1.52 & 6.2 & 6.4 & 98--113 \\
    136 & 25 & 4.64--7.49 & 5.7--5.9 & 5.9--6.1 & 49--64 \\
  \end{tabular}
  \caption{Separation distance, $h$ (nm), at which the net physical repulsion on an Al$_2$O$_3$ particle equals the interface force per particle applied in the simulations ($I = 10^{-4}$~M; each entry gives $h$ at the upper and at the lower end of the force range).}
  \label{tab:hrange}
  \end{center}
\end{table}

This demonstrates that the force magnitudes applied in the numerical simulations, 0.38--7.5~pN, are of the same order as the physical surface forces on these particles between about 5 and 30~nm from the interface. This supports treating $C_{int}$ as a lumped representation of those forces applied at the interface using the implementation described in Sec. \ref{sec:implementation}. The pH dependence shown here is also consistent with the assessment of varying $C_{int}$: the acid-stabilised suspension corresponds to the weakly repulsive end of the range, in which particles are permitted close to the interface (the model's small-$C_{int}$ behaviour), whereas suspensions above the isoelectric point correspond to a strong, longer-range repulsion that keeps particles away from the bubble surface (large $C_{int}$). There are, however, two limitations with this simplified implementation. The model force acts over the numerical interface thickness of 17--28~$\mu$m, so the comparison is one of magnitudes at the interface rather than of profiles (figure \ref{fig:netforce}\textit{a,b}); and the model does not consider an attractive component, so any near-interface accumulation predicted at acidic pH is represented only indirectly, through the weaker exclusion at small $C_{int}$.

\section{Extended Brownian force implementation details}\label{appC}
The Brownian force model used in the present solver was adapted from the compressible OpenFOAM particle library and made compatible with the incompressible, non-isothermal Eulerian-Lagrangian framework developed in this work. The objective of this modification was not to introduce a new physical model for Brownian motion, but to incorporate an established stochastic particle-force formulation into the present liquid-phase solver in a manner consistent with the governing equations described in section~\ref{sec:formulation}. A further correction was made to the statistical representation of the random forcing. Rather than applying a scalar random magnitude together with an imposed direction, the corrected implementation applies the Brownian contribution as a genuine three-dimensional Gaussian stochastic vector, consistent with equation~(\ref{eq:Brownian_force}). This modification was required to recover the correct second-order statistics of particle dispersion in isothermal conditions and to ensure that the non-isothermal drift arising from temperature-dependent Brownian forcing is represented consistently. A validation of this implementation has been presented in figure \ref{fig:validation_brownian_all}.

\section{Mesh sensitivity analysis}\label{appD}

The coupled Eulerian-Lagrangian formulation introduces two distinct mesh constraints. On the one hand, the Eulerian mesh must be sufficiently fine to resolve the near-wall temperature field and recover mesh-independent heat transfer predictions. On the other hand, excessive local refinement may affect the averaging of Lagrangian particle properties onto the Eulerian grid, particularly in the strongly biased near-wall cells. The mesh selection was therefore based on both the Eulerian thermal response and the minimum cell-size-to-particle-size ratio.

\begin{figure}
    \centering
    \includegraphics[width=1.0\textwidth]{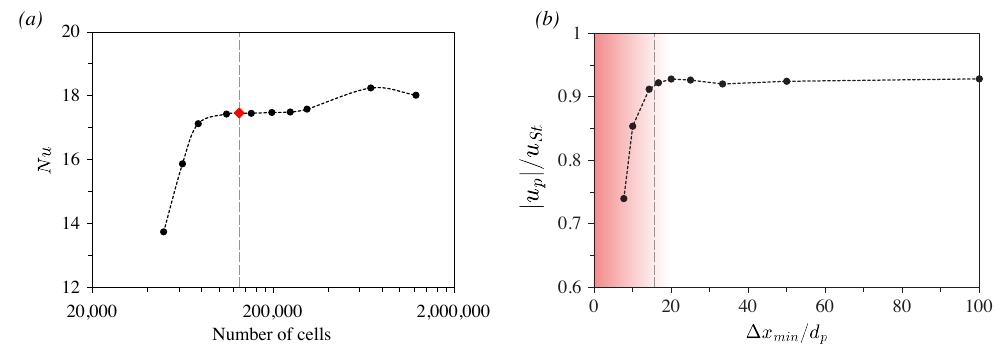}
    \caption{Cell size and mesh sensitivity assessment. \textit{(a)} Predicted wall-averaged Nusselt number for Taylor flow with $d_p = 100~\mathrm{nm}$, $\phi = 10^{-3}$, $Re = 1000$, and different computational grid sizes. \textit{(b)} Predicted normalised particle settling velocity using different cell-size-to-particle-diameter ratios.}
    \label{fig:mesh_sensitivity}
\end{figure}

The first assessment considered the wall-averaged Nusselt number for a particle-laden Taylor flow problem with \(Re=1000\), \(d_p=100~\mathrm{nm}\), and particle volume fraction, $\phi = 10^{-3}$. The predicted Nusselt number increased from 13.74 on the coarsest mesh (\(49\,352\) cells) to 17.45 at \(109\,802\) cells, demonstrating the sensitivity to the Eulerian resolution (figure~\ref{fig:mesh_sensitivity}\textit{a}). Beyond this level, however, the variation became much weaker; between \(129\,350\) and \(306\,383\) cells, the wall-averaged Nusselt number changed by less than 1\%. This indicates that the Eulerian predictions become effectively mesh independent once the mesh exceeds approximately \(1.1\times10^5\) cells.

A second assessment was then carried out using the normalised mean settling velocity, \(|u_p|/u_{St}\), for the same $d_p = 100~\mathrm{nm}$ particle dispersion considered in section~\ref{sec:ppp_sensitivity}. In this test, \(|u_p|/u_{St}\) was used to examine the effect of the minimum cell-size-to-particle-size ratio, \(\Delta x_{\min}/d_p\). The predicted settling response was found to increase rapidly as \(\Delta x_{\min}/d_p\) was raised from 7.7 to 16.7, but remained nearly constant for \(\Delta x_{\min}/d_p \ge 16.7\), with variations confined to within about \(\pm0.5\%\) (figure \ref{fig:mesh_sensitivity}\textit{b}). This behaviour suggests that excessive near-wall refinement can influence the particle averaging procedure when the smallest Eulerian cells become too small relative to the particle diameter $(\Delta{x_{min}}/d_p < 16.7)$.

The final mesh was therefore chosen to satisfy both numerical constraints simultaneously. A mesh of \(129\,350\) cells was adopted, corresponding to a minimum cell size of \(3.65~\mathrm{\mu m}\) for a $H = 1~\mathrm{mm}$ channel. For the particle-diameter range considered in the present study (\(78\) to \(136~\mathrm{nm}\)), this gives a minimum cell-size-to-particle-size ratio between 26.8 and 46.8, which lies safely within the plateau region identified in figure \ref{fig:mesh_sensitivity}\textit{b}. In addition, consistent with the criterion adopted in our previous work on non-Brownian and isothermal Taylor flow suspensions \citep{Khaleel2025}, the liquid film between the bubble and the wall was always resolved by more than five cells. The adopted mesh therefore satisfied the lower resolution limit required by the Eulerian fields, while remaining sufficiently coarse near the wall to preserve a stable Lagrangian particle response.

\bibliographystyle{unsrtnat}
\bibliography{jfm}

\end{document}